\documentclass[journal]{IEEEtran}

\usepackage[T1]{fontenc}
\usepackage{tabularx}
\usepackage{cite}
\usepackage{graphicx}
\usepackage{amsmath}
\usepackage[cmintegrals]{newtxmath}
\usepackage{algorithmic}
\usepackage{subcaption}
\usepackage{xcolor}
\usepackage[para]{footmisc}
\usepackage{tabularx,multirow,array}
\usepackage{multirow}
\usepackage{booktabs}
\usepackage[linesnumbered,ruled,vlined]{algorithm2e}
\SetKwInput{KwInput}{Input}                
\SetKwInput{KwOutput}{Output}  
\SetKwInput{KwReturn}{Return}  
\usepackage{textcomp}
\usepackage{linguex}
\usepackage[linguistics,edges]{forest}

\usepackage[shortlabels]{enumitem}
\usepackage{tikz}
\usepackage{tikz-qtree}
\usetikzlibrary{trees} 
\usepackage{array}
\newcommand{\readded}

\usepackage{metalogo}
\usepackage{smartdiagram}
\usepackage{algorithmic}
\usepackage{tabularx,multirow,array}
\usepackage{url}
\usepackage{csquotes} 
\usepackage{arydshln} 
\usepackage{color, soul}
\usepackage{makecell}
\usepackage{cite}
\usepackage{etoolbox}
\usepackage{threeparttable}
\usepackage{pifont}
\usepackage{dblfloatfix}
\usepackage{float}
\usepackage[normalem]{ulem}

\usepackage{amsfonts}
\usepackage{mathtools}
\DeclarePairedDelimiter{\ceil}{\lceil}{\rceil}
\RequirePackage{tikz}
\RequirePackage{etoolbox}
\RequirePackage{xparse}
\RequirePackage{xstring}
\newcommand{\etal}{\textit{et al. }}
\newcolumntype{P}[1]{>{\centering\arraybackslash}p{#1}}
\usepackage{balance}

\begin{document}
\title{DRLD-SP: A Deep Reinforcement Learning-based Dynamic Service Placement in Edge-Enabled Internet of Vehicles}

\author{Anum~Talpur,~\IEEEmembership{Member,~IEEE,}
	and~Mohan~Gurusamy,~\IEEEmembership{Senior~Member,~IEEE}
	\thanks{A. Talpur and M. Gurusamy are with the Department of Electrical and Computer Engineering, National University of Singapore, Singapore (email: anum.talpur@u.nus.edu; gmohan@nus.edu.sg).}}


\maketitle

\begin{abstract}
The growth of 5G and edge computing has enabled the emergence of Internet of Vehicles. It supports different types of services with different resource and service requirements. However, limited resources at the edge, high mobility of vehicles, increasing demand, and dynamicity in service request-types have made service placement a challenging task. A typical static placement solution is not effective as it does not consider the traffic mobility and service dynamics. Handling dynamics in IoV for service placement is an important and challenging problem which is the primary focus of our work in this paper. We propose a Deep Reinforcement Learning-based Dynamic Service Placement (DRLD-SP) framework with the objective of minimizing the maximum edge resource usage and service delay while considering the vehicle's mobility, varying demand, and dynamics in the requests for different types of services. We use SUMO and MATLAB to carry out simulation experiments. The experimental results show that the proposed DRLD-SP approach is effective and outperforms other static and dynamic placement approaches.
\end{abstract}


\IEEEpeerreviewmaketitle

\section{Introduction}
\IEEEPARstart{T}{he} evolution of fifth-generation network (5G) brings enormous benefits in Internet of Vehicle (IoV) networks. It contributes to intelligent and sustainable vehicular networks with advanced safety, reliability, transportation efficiency, low latency, and wider network coverage \cite{IoV5G}. 5G networks are end-to-end programmable networks that provide quality performance while meeting the requirements of multiple services. The next-generation mobile network (NGMN) association has proposed the concept of network slicing\cite{slicing}, where network slices are virtual network functions over a common physical network to satisfy different service-requirements. The International Telecommunications Union (ITU) has classified different 5G services into three major application scenarios, namely, enhanced Mobile Broadband (eMBB), ultra-Reliable and Low Latency Communications (URLLC), and massive Machine Type Communications (mMTC) \cite{ITUSlice}. These applications provide high data rates, high reliability, low latency, and high connection density. To support parallel functioning of multiple applications, several computational operations need to be performed within a network. The European Telecommunications Standards Institute (ETSI) introduces the use of mobile edge computing (MEC) with IoV networks which extends storage and compute resources of cloud bringing them closer to the end user \cite{MECSurvey}. It provides better coverage for vehicles and fulfills various service requirements including, high reliability, low latency, security, and so on\cite{etsiMEC}. \par
\begin{figure}[htbp]
	\begin{center}
		\includegraphics[width=3.1in,height=1.9in]{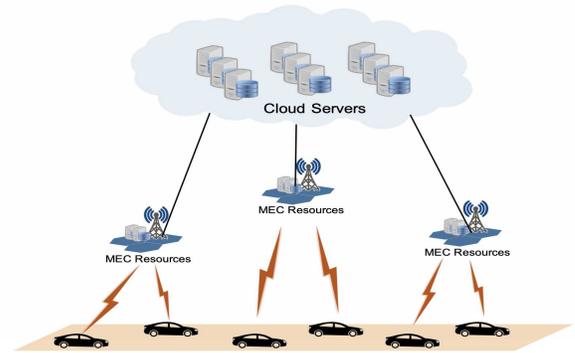}
		\caption{The framework of three-layer IoV Network}
		\label{fig:Intro}
	\end{center}
\end{figure}

  
Fig. \ref{fig:Intro} shows a framework of a three-layer IoV network where vehicles communicate with the infrastructure for services like media downloading, cooperative awareness message (CAM), decentralized environmental notification messages (DENM) and so on, to avail coordination in remote driving, parking space discovery, navigation, road safety, and many other applications. Multiple services can be deployed at the edge servers making use of compute, network and storage resources. One of the primary challenges in IoVs is service placement. Service placement is the problem of mapping services to the edge servers in IoVs to satisfy the requirements for the requested services while using the edge resources efficiently. From the user perspective, it is important to minimize the delay perceived by a vehicle. From the service provider's perspective, edge resource usage is an important metric that should be minimized while keeping the resource load across the servers as balanced as possible. This will enable servers to scale up the resources for varying future demands and efficiently handle the events of congestion and failures. \par
The growing complexity of traffic patterns and dynamics in the requests for different types of services has made service placement more challenging. It is necessary to adopt continual learning of the environment for providing better services. Embedding intelligence using machine learning (ML) has drawn research interests recently. Different ML techniques have received significant attention and reinforcement learning (RL) is an attractive approach for various problems in the area of vehicular communications \cite{RLforITS}. Considering the mobile and changing environment of IoVs, the reinforcement learning algorithm has the capability to train models online as the system operates without prior knowledge. Q-learning and deep reinforcement learning (DRL) are promising in many edge-enabled IoV applications such as motion planning \cite{driving}, resource sharing/caching schemes \cite{RL1,cache3,cache2}, offloading \cite{R1,R2}, scheduling \cite{schedule,schedule2}, navigation \cite{navigation} and so on. A reinforcement learning framework contains an agent that interacts with the environment to observe the state, take an action, and in response receives a reward/punishment to enhance the performance of the network for future actions. The objective is to learn a policy which maximizes the reward or minimizes the punishment, respectively. The actor-critic approaches of DRL have been widely explored in the literature to deal with continuous control problems \cite{RLforITS}. In our work, we extend the traditional actor-critic framework along with integer linear programming (ILP) formulation to solve the service placement problem. Specifically, the DRL agent uses actor network as policy function and critic network as value-function to design deep reinforcement learning-based dynamic service placement (DRLD-SP) framework. Our DRLD-SP framework leverages the ILP-based optimization formulation as an actor network to yield much faster learning for service placement. On the other hand, the critic network uses a deep neural network (DNN) to train its network to quickly specify quality values for decisions taken by policy function (actor network). \par
In our previous work in \cite{Anum}, we addressed the problem of service placement in vehicular networks using delay or edge resource usage as the objective function. We proposed a reinforcement learning-based (Q-learning) dynamic service placement framework to find the optimal placement of services at the edge servers while considering the vehicle's mobility and dynamics in the requests for different types of services. Our work in \cite{Anum} doesn't consider the increasing demand from vehicles and only performs one-to-one placement (i.e placement of only one service at one edge node). In addition, it maintains a look-up table (Q-table) to keep a record of quality decisions. Different from \cite{Anum}, in this work, we propose deep reinforcement learning based on-demand DRLD-SP system of many-to-one placement where the number of service instances varies based on the requirement from vehicles. It has the ability to scale up or scale down the usage of resources at the edge with changing service demands. This helps to keep a balance of resources (from a service provider perspective) to efficiently handle the events of congestion, failures, and varying traffic conditions while satisfying the adequate delay from the perspective of vehicles. Considering this, we propose a single objective function that minimizes the maximum of both edge resource usage and service delay, and controls the relative importance of resource usage vs. service delay by using a parameter $\alpha$. Moreover, our DRLD-SP framework proposes a new solution approach to evaluate decision quality value by using a neural network. Q-learning is not efficient for growing and complex IoV networks to store all quality values in one table. It is also time-consuming to perform a frequent query in a large table. Therefore, in this work, we adopt deep reinforcement learning to overcome the limitation of Q-learning in terms of value-function approximation ability. We use SUMO and MATLAB to carry out simulation experiments. The main contributions of our DRLD-SP framework are as follows:
\begin{itemize}
	\item We consider a three-layer edge-enabled IoV network and formulate the dynamic service placement problem with the goal of minimizing the maximum edge resource usage (from the service provider's perspective) and service delay (from a user perspective).
	\item We propose DRLD-SP agent which consists of policy function (actor network) and value-function (critic network). The actor network uses an ILP formulation to make service placement decisions, whereas the critic network uses DNN to evaluate the performance of decisions taken by the actor network in terms of delay observed by vehicles. 
	\item Performance evaluation is carried out on realistic IoV traces created virtually using SUMO simulator. The results show that DRLD-SP outperforms the exisitng static and dynamic placement schemes.
\end{itemize}
The rest of the paper is organized as follows. Section \ref{sec:relatedwork} provides an overview of the related work in the literature. Section \ref{Sec:System-Model} describes the system architecture, network and service request model, computing model, placement problem and proposed approach. Section \ref{sec:DRLD} describes the proposed method. Section \ref{sec:results} discusses the experimental setup, simulation metrics and results. Section \ref{sec:conclude} makes concluding remarks.

\section{Related Work}
\label{sec:relatedwork}
The problem of service placement in IoV networks is not widely explored in the literature. Some recent works \cite{Edge-enabledV2X,multicomponent-V2X,costoptimalplacement} study the static service placement problem and develop solutions to produce a fixed mapping of services to edge servers for a problem scenario. In a recent work \cite{Edge-enabledV2X}, the authors consider the problem of vehicle-to-everything (V2X) service placement. They propose an ILP model for minimizing the average service delay. The scope of the experiments is limited to highway environments where the speed of vehicles is constant with a fixed distance between vehicles, and the movement of vehicles in one direction. The delay obtained for V2X communication is also on randomly assigned values from a given set of ranges. This work does not study the changing traffic patterns and time-varying nature of vehicular environment while making service placement decision. In \cite{multicomponent-V2X}, the authors consider a highway environment for V2X service placement. They propose a binary ILP model for minimization of communication and download link delay for five different V2X applications. In \cite{costoptimalplacement}, the authors present cost-focused delay-aware V2X service placement. It also considers one-time service placement and doesn't encounter the changing environment of vehicles. Moreover, for the physical environment, the authors consider a highway scenario of 2 lanes with the delay observed by vehicles is estimated using the uniform distribution for a given range of values. Some work on V2X applications is carried out in the context of cloud computing and fog-computing \cite{cloud2,cloud3,priority1}. One common aspect in most of the previous works is the static placement (i.e. one-time placement) and consideration of latency or delay as the objective. In few works, the service-type priority \cite{priority1} and cost \cite{cloud2,cloud3,costoptimalplacement} are used as an additional factors for service migration. \par
In the literature, dynamic algorithms are proposed which address the mobility features of users \cite{D1latencyaware,D2drl,D3cloud,D4fog}. In a recent work\cite{D1latencyaware}, Mada \etal propose service migration scheme for 5G mobile systems. This work proposes an ILP formulation with the objective of minimizing resource allocation while migrating services across centralized cloud and edge cloud. To cope with the varying mobility patterns of users, this work proposes to always reoptimize during an epoch without any prior knowledge on the need for service migration. Here, the user requests and delay observed are randomly chosen from a given range of values. This work does not consider the real mobility patterns of vehicles. The migration cost is also not taken into consideration in this work, which is generally high for such "always migration" schemes. The works in \cite{D3cloud,D4fog} propose to use threshold-based migrations in wireless networks. Here, the state transition conditions are specified and whenever the parameter (such as distance, number of hops or RSSI value) exceeds the threshold value, the service migration scheme is automatically triggered. In such a scheme, the selection of a threshold value is complex and requires careful consideration and complicated theoretical analysis. The use of a fixed threshold for vehicular networks is not a good choice where service requirement parameters widely differ between safety applications and value-added applications (e.g. infotainment). A recent work in \cite{D2drl} proposes DRL-based service migration in vehicular networks. Different from our work, the focus of this work is on migration and migration frequency, and triggers service migration decisions by considering the velocity of vehicles. However, this work may not satisfy delay requirements for the requests which is an important requirement for many vehicular services. In addition, the dynamicity in vehicular networks is not only due to the mobility of vehicles. The above works fail to consider the dynamicity in terms of service requests and increasing/decreasing demands from users. We propose to use DRL to address the varying traffic patterns as well as dynamicity of service requests. We choose to use reinforcement learning because it's scalable, considers infinite state space, has the ability to interact with environment and address the changing conditions to make dynamic decisions. \par    
The solutions have been developed in literature using DRL for many other problem statements. Liang \etal in \cite{dynamicRA} propose a Q-learning based dynamic resource allocation mechanism for services. The objective of this work is to maximize the system's computation revenue and minimize the service rejection rate. This work performs priority-based allocation for services. Here, the arrival rates are randomly chosen and do not encounter the real-time mobility of vehicles. Wang \etal in \cite{RL1} propose a DRL-based resource allocation mechanism for edge nodes in vehicular networks. This work focuses on the resource sharing scheme for edge nodes where nodes cooperate for optimizing resources for different tasks. Their design doesn't evaluate realistic traffic scenarios. The incoming traffic and delays are modeled with queuing theory using the probability distribution function. The use of DRL is also observed in the area of content caching and content sharing. Qiao \etal in \cite{cache3} propose a framework for cooperative content caching between vehicles and RSUs. The authors use cost (i.e. price per resource unit) as their objective and satisfying delivery latency as the constraint. Using travel history as an input to a DRL framework, the authors present an algorithm to perform a joint-caching decision. Another DRL-based work on cost-efficient joint optimal caching is carried out in \cite{cache2}. This work uses deep Q-learning to estimate the set of possible connecting RSUs and vehicles for caching placement. Different from the problem we focus in this paper, in the above works, vehicles generate contents about road status, driving patterns, sensing information, and so on, and offload to other vehicles and infrastructure for further processing and sharing. Such contents have different data sizes and computation demands, depending on the vehicles generating them. \par
Recently, some researchers have studied the use of DRL in computation offloading  and scheduling for vehicular applications \cite{schedule,R1,R2}. Wang \etal in use DRL for scheduling and decision making regarding the charging and relocation recommendation system of e-taxis. It uses a non-cooperative DRL framework in vehicle's ride-hailing platform to decide on charging their batteries or serving order first. It helps vehicles to avoid the situation of battery running out of the charge. In \cite{schedule}, Zhan \etal propose a DRL-based computation offloading scheduling scheme where vehicles traveling on the expressway schedules the waiting tasks in the queue to minimize latency and energy consumption. Another work on computation offloading \cite{R1} presents an optimal solution for calculating offloading proportion. This work considers a single-user scenario and assumes that the task can be decomposed into several subtasks, which can be executed in parallel across multiple nodes. Ning \etal \cite{R2} present a supervised learning-based computation offloading and content-caching algorithm. It trains a binary classifier using SVM based on the data collected and decisions made from the proposed formulation, to choose the best node for offloading. Different from our work, in computation offloading, the vehicle is the initiator to upload tasks or part of a task (after decomposing a task into several subtasks) to other vehicles or edge nodes for availing computation resources. Not limited to DRL, the use of graph theory is also recently explored for scheduling applications because of its small scale inputs in terms of the number of tasks \cite{containerization}. However, it may not be a good choice for service placement applications where input is of large scale (i.e vehicle dynamics). \par 
Our work considers the dynamic service placement problem and develops a DRLD-SP framework to handle the dynamicity of vehicles considering dynamic user demands and vehicle mobility. We use edge resource usage (from service provider perspective) and service delay (from user perspective) as important metrics to optimize. The number of deployed service instances also varies (increases or decreases) based on the varying demand from vehicles to maintain efficient usage of limited edge resources. 

\section{System Description and Problem Statement}
\label{Sec:System-Model}
This section provides an overview of the hierarchical architecture of the IoV system. Then, the network and service request model, and computing model are discussed. Finally, we describe the service placement problem and our proposed approach.

\subsection{System Architecture}
\begin{figure}[htbp]
	\begin{center}
		\includegraphics[width=3.3in,height=2.25in]{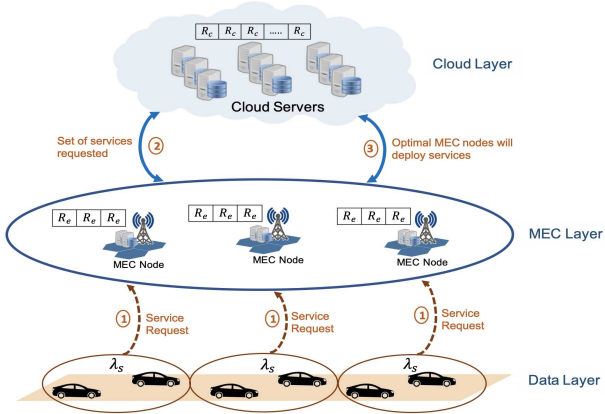}
		\caption{Hierarchical IoV architecture for our proposed service placement approach}
		\label{fig:Architecture}
	\end{center}
\end{figure}
The hierarchical architecture of our proposed IoV system is depicted in Fig. \ref{fig:Architecture}. It consists of three layers that include data layer, MEC layer, and cloud layer. Here, MEC extends the capabilities (storage and compute) of cloud and brings them closer to the end user. At the data layer, we assume a city road environment with multiple lanes and movement of vehicles in different directions. The vehicles are randomly choosing a source, destination, and speed to start and end their journey at different times. The speed limit regulations and vehicle arrival rates specified in the SUMO simulator, for the type of vehicle and environment are followed. The service request is generated from the data layer with a uniformly distributed arrival rate $\lambda_s$ for the type of service $s$. We assume the IoV network environment is under 5G coverage using evolved NodeB (eNB) stations for which inter-site distance (ISD) is 500m (urban-macro 5G regulations) \cite{etsi5G}. It is also assumed that there are multiple eNBs equipped with MEC hosts to form the network edge with limited capacity servers. The available resource capacity at edge $C_e$ which includes compute, network and storage resources, are calculated as $C_e=\sum_{n=1}^{N}R_e(n)$. Here, $R_e$ denotes a virtual resource unit and $N$ denotes the total number of resource units at the edge. Additionally, the network edge connects to large capacity cloud servers (at the cloud layer) via a backbone network. The resources at cloud $C_c=\sum_{m=1}^{M}R_c(m)$ are larger where $M >\!\!>\!\!> N$. Here, $R_c$ denotes a virtual resource unit and $M$ denotes the total number of resource units at the cloud. Due to MEC capacity limitations, the placement of services at the edge only takes place when there is a demand for that service. In case of no demand, the edge node will remove the instance of service $s$ from its resources and transfer it back to the cloud to keep MEC nodes less loaded and give better performance for new service demands. We do not consider the communication channel characteristics, and we assume adequate links between different layers, nodes, and servers are available to enable communication among them.\par

\subsection{Network and Service Request Model}
\label{Sec:NetworkANDserviceRequestModel}
We use $E$ to denote a set of edge servers with $e$ $\in$ $E$ as an edge node. For each edge node $e$, the residual resource capacity (available resources) is denoted by $C_e$. Let $V$ and $S$ denote a set of vehicles and service types (services), respectively. A vehicle $v \in V$ requires a service $s \in S$ which is to be hosted at a MEC node. The number of vehicles requesting service $s$ is denoted as $\lambda_s$, and the number of vehicles one instance of service $s$ can handle (or provide parallel connection) at a time is denoted by $\mathbb{C}$. Further, a service request model is defined as a 4-tuple structure ($\nu$, $loc$, $t$, $s$). We assume each vehicle $\nu$ is equipped with a clock and GPS, which enable it to specify time $t$ and location $loc$ in its service request message. Associated with each service $s$, the amount of resources consumed by deploying it at edge node $E$ is denoted by $R_s$, and the delay/latency requirement threshold is denoted as $D_s$. In response to the request, the location of optimal MEC nodes/servers will be calculated to deploy services. The notations are summarized in Table. \ref{tab:notations}.
\begin{table}[htbp]
	\scriptsize
	\centering
	\caption{Summary of Notations}
	\begin{tabular}{lp{6.5cm}}
		\toprule
		\textbf{Notation} & \textbf{Description} \\
		\midrule
		$E$    & Set of MEC nodes \\
		$S$     & Set of services \\
		$V$     & Set of vehicles \\
		$R_s$    & Resources consumed by service $s$ \\
		$C_e$ & Available resources at edge node $e$ \\
		$x_e^s$ & Assignment of service $s$ at the edge node $e$ \\
		$D_s$ & Delay threshold or maximum allowed delay for service $s$  \\
		$\lambda_s$    & Number of vehicles requesting service $s$ \\
		$\mathbb{C}$ & Number of vehicles a service instance can handle at a time\\
		$\mathcal{I}_s$ & Number of instances required for service $s$\\		
		$\varphi_e$ &  The edge resource usage \\
		$d^s_{e}$  & The average time delay experienced by vehicles when service $s$ is deployed at node $e$ \\
		\bottomrule
	\end{tabular}%
	\label{tab:notations}%
\end{table}%
\subsection{Computing Model}
\label{section:computingModel}
We model the MEC computation system as M/D/1 queue, where arrival occurs with $\lambda_s$ according to Markov stochastic model and service processing rate is deterministic (serving at rate $\mathbb{C}$). The total service delay observed by vehicles while requesting for service $s$ from edge node $e$ refers to the total time from when a vehicle sends a service request $s$ to when the corresponding response is received from an edge node. In our proposed computation model, it consists of propagation delay and queuing delay: $d_{e}^s=d_{prop}^s+d_{queue}^s$. We assume that there is no waiting queue if $\lambda_s \le \mathbb{C}$. In such cases, the queuing delay will become zero i.e. $d_{queue}^s=0$. However, if $\lambda_s>\mathbb{C}$, a queue will be created for vehicles more than $\mathbb{C}$ and the average waiting time for service $s$ over the MEC node will be calculated as \cite{queueMD1}:
\begin{equation}
d_{queue}^s=\frac{\grave{\lambda_s}}{2\mathbb{C}(\mathbb{C}-\grave{\lambda_s})} 
\end{equation}
Here, $\grave{\lambda_s}=\lambda_s-\mathbb{C}$. As a rule of thumb, the average propagation delay is computed as the ratio between the distance and the propagation speed over the medium:
\begin{equation}
d_{prop}^s=\frac{1}{|V|}\sum_{v\in V}\frac{dist(v,s)}{c}
\end{equation}
Here, $dist(v,s)$ is the euclidean distance between vehicle $v$ and the node where service $s$ is deployed, and $c$ is the propagation speed of the signal through communication medium. Thus, the total service time can be obtained by:
\begin{equation}
d_{e}^s= \frac{1}{|V|}\sum_{v\in V}\frac{dist(v,s)}{c} + \frac{\grave{\lambda_s}}{2\mathbb{C}(\mathbb{C}-\grave{\lambda_s})} 
\label{eq:delay}
\end{equation}
For analyzing the load over a MEC node, we calculate the edge resource usage which is denoted by $\varphi_e$. It is a ratio between the resources that $\mathcal{I}_s$ instances of service $s$ will consume and the available resources at the edge node. We can calculate it as:  \par
\begin{equation}
\varphi_e = \frac{\sum_{\mathcal{I}_s} R_s }{C_e}, \forall e\in E, \forall s\in S,
\label{eq:resouceUsage}
\end{equation}
The calculation of  $\mathcal{I}_s$ is based on $\mathbb{C}$ and given by:
\begin{equation}
\mathcal{I}_s=\ceil[\bigg]{\frac{\lambda_s}{\mathbb{C}}}, \forall s\in S
\label{eq:instance}
\end{equation}
We extract the features of service requests and use deep reinforcement learning to solve the service placement problem in an on-demand dynamic manner. 

\subsection{Service Placement Problem and Proposed Approach}
We consider service placement problem in IoV networks with MEC nodes having limited resources. Given a set of services with their resource and delay requirements, the problem is to find the optimal placement of services at the edge servers while considering the vehicle's mobility and dynamics in the requests for different types of services. The number of vehicles requesting service $s$ and their distance from different edge servers are dynamic. A static solution (SSP) which fixes servers for hosting services is not effective for a mobile and dynamic scenario of an IoV network. It is therefore imperative that the real-time environment be taken into consideration while mapping a service to an edge server. \par
With this goal, we proposed an RL-based dynamic service placement approach in \cite{Anum}. The work in \cite{Anum} uses a classic model-free Q-learning algorithm that optimizes a certain objective such as minimizing resource usage or minimizing the delay. In this work, we proposed a single objective function that minimizes the maximum of both edge resource usage and service delay, and controls the relative importance of resource usage vs. service delay by using a parameter $\alpha$. In \cite{Anum}, we considered a one-to-one placement with a fixed number of services and availability of a single instance for each service. However, deployment of one instance for one service can only provide service to a limited number of vehicles. In case of an increase in demand for any particular service, there will be a need for multiple instances for each service. In this work, we propose an on-demand system of many-to-one placement where the number of service instances varies based on the requirement from vehicles. It has the ability to scale up or scale down the usage of resources at the edge with changing service demands. This helps to keep a balance of resources (from service provider perspective) to efficiently handle the events of congestion, failures, and varying traffic conditions while satisfying the adequate delay from the perspective of vehicles. Different from our work in \cite{Anum}, we propose a DRLD-SP framework in this work. Q-learning used in \cite{Anum} has scalability problem with large tables for complex IoV networks. Therefore, a deep learning model will help to provide a quick solution (remapping of services) by estimating the quality of performance metrics which will mitigate the poor performance at any particular communication link or channel between the vehicle and server.

\section{DRLD-SP: Deep Reinforcement Learning-based Dynamic Service Placement}
\label{sec:DRLD}
\begin{figure}[htbp]
	\begin{center}
		\includegraphics[width=2.55in,height=1.65in]{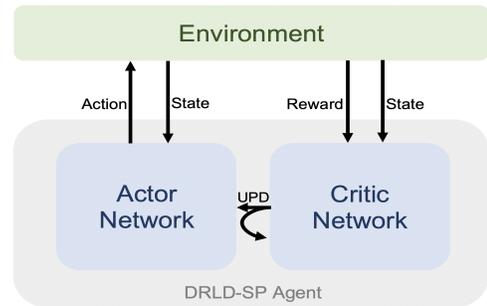}
		\caption{Structure of the DRLD-SP Agent}
		\label{fig:Agent}
	\end{center}
\end{figure}

In this section, we present the proposed DRLD-SP framework, for the problem described above. We exploit the actor-critic DRL model with an ILP formulation to solve the service placement problem in a mobile scenario of IoV networks. The block diagram of the actor-critic DRLD-SP agent is shown in Fig. \ref{fig:Agent}. The DRLD-SP agent learns and updates the actor-critic networks by interacting with the time-varying IoV environment. An actor generates action and a critic estimates a value-function needed to keep the performance of an actor updated. We leverage the actor-critic with our ILP formulation to perform optimal service placement in a dynamic manner. \par
\begin{figure*}[htbp]
	\begin{center}
		\includegraphics[width=5.2in,height=3in]{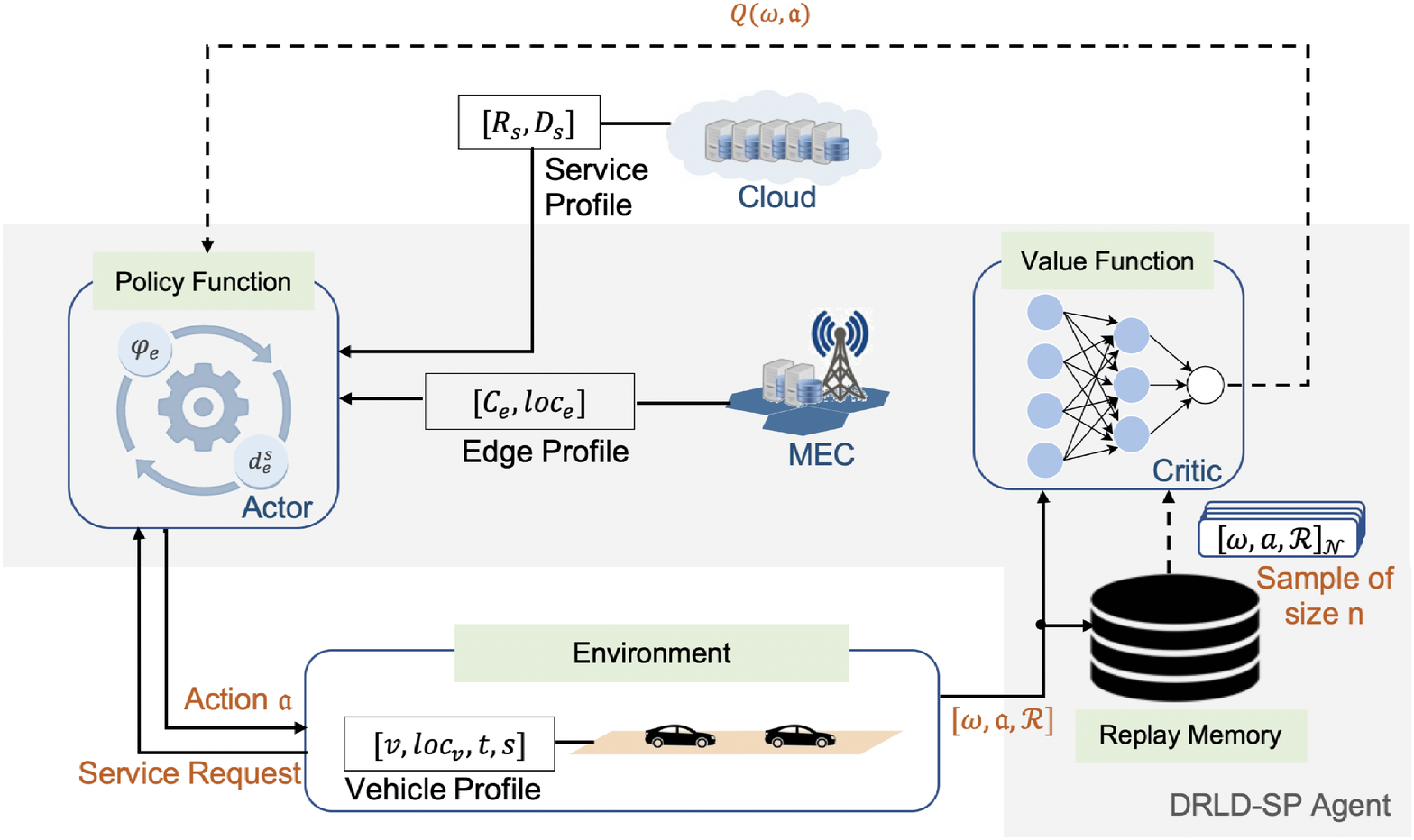}
		\caption{The proposed DRLD-SP approach }
		\label{fig:Model}
	\end{center}
\end{figure*}
First, we briefly explain the design of state space, action space, policy and reward function used in our DRLD-SP framework. 
\subsection{State Space $\omega$}
\label{Sec:state}
At a given time instant $t$, the state space set describes the network environment. The DRLD-SP agent observes an environment to constitute the following set of data $\omega$ from the service request model:
\begin{equation}
\omega=\{[v_1,loc_{1}, s], [v_2,loc_{2}, s], ... , [v_n,loc_{n}, s]\}_t
\label{eq:omega}
\end{equation}
where $s\in S$, {$v_1, v_2, ...,v_n$} is the set of vehicles IDs, and {$loc_1, loc_2, ..., loc_n$} is the set of locations of vehicles requesting for service $s$ at time unit $t$. 
\subsection{Action Space $\mathfrak{a}$}
The action space describes the action taken by the policy module for placement of service $s$ on edge node $e$, as shown in Fig. \ref{fig:Model}. Let $\mathfrak{a}$ denote the action space. The action at time unit $t$ is defined as:
\begin{equation}
\mathfrak{a}=\pi(\omega)=x_e^s, \forall e\in E, \forall s\in S,
\end{equation}
where $\pi$ is the policy (defined in the next section) required to generate an action over the observation set of $\omega$ at time unit $t$, and $x_e^s$ gives the matrix indicating the placement of service $s$ on edge node $e$.
\subsection{Policy Function $\pi$}
\label{Sec:policy}
The policy $\pi$ is a function performed by an actor network to map state-space to an action-space $\pi: \omega \rightarrow \mathfrak{a}$. For DRLD-SP, the actor network performs a policy to optimize the objective function subject to different constraints, as shown in Fig. \ref{fig:Model}. We use a single objective function in the framework of our study. The objective is to minimize the maximum edge resource usage and service delay, and control the relative importance of resource usage vs. service delay by using a parameter $\alpha$. The rationale for using resource usage is to efficiently utilize the limited edge resources and decrease the possibility of congestion so that the MEC node has enough room for service instance scale-up in case of increased future demands. From the perspective of a user, minimizing the maximum delay will help to satisfy adequate delay requirements and make service availability faster for the vehicles. The policy function $\pi$  is formulated as:
\begin{equation}
\pi = \underset{s \in S,e \in E}{minmax} \hspace*{3mm} \left(\alpha \sum_{e\in E} \varphi_e + (1-\alpha) d_{e}^s x_e^s\right)
\label{eqObj1}
\end{equation}

The objective of the problem is to minimize the maximum edge resource usage and total service delay observed by vehicles. The edge resource usage $\varphi_e$ is determined as the ratio between the resources that $\mathcal{I}_s$ instances of service $s$ will consume and available resources at edge node $e$, as described in Section \ref{section:computingModel}. Whereas the service delay $d_{e}^s$ consists of propagation delay and queuing delay observed by a set of vehicles while requesting for service $s$ from edge node $e$. The queuing delay is obtained through approximating edge computation system as M/D/1 system, as described in Section \ref{section:computingModel}. Note that the service delay is normalized in the range [0,1] by diving it to the maximum possible service delay. We introduce a parameter, $\alpha$, to control the relative importance of resource usage vs. service delay. The placement of service $s$ at edge node $e$ is given by $x_e^s$, where $x_e^s$ is a binary variable. If edge node $e$ deploys service $s$, $x_e^s$ is 1. Otherwise, it is 0. The placement of service is subjective to the following constraints: \par
\begin{description}
	\item \textbf{Mapping Constraint:} 
	This constraint guarantees each edge server node hosts a service or a set of services, and the decision variable $x_e^s$ is a binary integer decision variable.
	\begin{equation}
	\sum_{s\epsilon S}x_e^s \ge 1; \forall e\in E
	\label{eqC1}
	\end{equation} 
	\\where,
	\begin{equation*}
	x_e^s \in \{0,1\}; \forall s\in S, \forall i\in E
	\label{eqx}
	\end{equation*} 
	\item \textbf{Delay Contraint:} 
	This constraint ensures that the service delay experienced by vehicles requesting service $s$ should be less than the service's maximum delay threshold $D_s$. 
	\begin{equation}
	\sum_{s\in S}x_e^s d_{e}^s \le D_s; \forall e\in E
	\label{eqC2}
	\end{equation} 
	\item \textbf{Resource Constraint:} 
	This constraint ensures that the available resources at the edge node are not exhausted while deploying $\mathcal{I}_s$ instances of service $s$, where $\mathcal{I}_s\ge 1$.
	\begin{equation}
	\sum_{e\in E}x_e^s\mathcal{I}_sR_s \le C_e; \forall s\in S
	\label{eqC3}
	\end{equation} 
\end{description}

\subsection{Reward $\mathcal{R}(\omega,\mathfrak{a})$}
\label{Sec:reward}
At each time unit $t$, in the response of the action taken by an actor network of the DRLD-SP agent, the system receives an immediate reward $\mathcal{R}(\omega,\mathfrak{a})$ from the environment. Generally, the DRL agent aims to maximize the reward. However, the objective of our service placement problem is to minimize the service delay observed from vehicles in accessing service $s$ from the associated edge server $e$. Therefore, the reward function is calculated as:
\begin{equation}
\mathcal{R}(\omega,\mathfrak{a})=\mathbb{E}\left[d_e^s(t)\right]
\label{eq:reward}
\end{equation}
where $d_e^s(t)=\frac{1}{|V|}\sum_{v\in V}\frac{dist(v,s)}{c} + \frac{\grave{\lambda_s}}{2\mathbb{C}(\mathbb{C}-\grave{\lambda_s})}$, is the average service delay observed by a set of vehicles at time unit $t$. 

\subsection{DRLD-SP Agent}
Fig. \ref{fig:Model} depicts the framework of the DRLD-SP algorithm which consists of \textbf{\textit{environment}}, \textbf{\textit{policy function}} (actor), \textbf{\textit{value function}} (critic), and \textbf{\textit{replay memory }}$\mathcal{M}$. The grey shaded area represents the computations or functions performed by \textbf{\textit{DRLD-SP agent}} over the MEC node. The actor network and critic network are the agent's primary functions to perform action and evaluate decision quality value. The DRLD-SP agent has direct interaction with the environment. \par 

From \textbf{\textit{environment}}, the request for service $s$ is initiated by vehicle $v$ following the service request model, as discussed in Section \ref{Sec:NetworkANDserviceRequestModel}. In return, considering the demand for service $s$ at time $t$ and location $loc$ of vehicles requesting for service $s$, \textbf{\textit{the policy function module}} selects the edge servers for the services for placement based on the action selection strategy $\pi$, as discussed in Section \ref{Sec:policy}. The task of \textbf{\textit{the value function module}} is to critic the performance of the actor network based on the action taken and rewards received. It is responsible for calculating the quality value $Q(\omega,\mathfrak{a})$ of the decision taken by the actor network of the policy module. A high $Q(\omega,\mathfrak{a})$ means a high-quality decision. Therefore, an actor has to select actions with the maximum quality value, $\mathfrak{a}=$ arg max $Q(\omega, \mathfrak{a})$. In our proposed design, the critic network is a neural network. The input of the neural network is a state, action, and reward. The reward is a response, an agent receives by the environment for the corresponding action, as discussed in Section \ref{Sec:reward}. The critic network updates its parameters $\theta$ to minimize the mean square loss function $\mathcal{L}_Q$. The loss function is computed as:

\begin{equation}
\mathcal{L}_Q(\theta)=\frac{1}{\mathcal{N}}\sum_{i=1}^{\mathcal{N}}\left[(y_{t_i}-Q_i(\omega,\mathfrak{a};\theta))^2\right]
\label{eq:loss}
\end{equation}
Here, $y_t$ is a target value which is calculated as:
\begin{equation}
y_t =
\begin{cases}
\sigma(D_s,\mathcal{R}(\omega,\mathfrak{a})) & \mathcal{R}(\omega,\mathfrak{a}) < D_s\\
0 & \text{else}
\end{cases} 
\label{eq:yt}
\end{equation}
Where $\sigma(D_s,\mathcal{R}(\omega,\mathfrak{a}))$ is the standard deviation between delay threshold and reward. The higher the deviation is, the better the model in terms of delay. DRLD-SP agent further uses a \textbf{\textit{replay memory}} $\mathcal{M}$. It is used to store the experience for training the critic network. The transition information contains $[\omega,\mathfrak{a},\mathcal{R}(\omega,\mathfrak{a}))]$, required to train a network. The critic network uses replay memory to fetch experience after a random period of time $T$ and optimizes the network parameters for better performance. \par
We present the proposed DRLD-SP agent framework in Algorithm \ref{Alg:DRLD} and Algorithm \ref{Alg:DRLD-Decision}. In Algorithm \ref{Alg:DRLD}, we present the network optimization and training procedure. In this algorithm, $\mathbb{U}$ is the total number of episodes required to train critic DNN, and $T$ is the time step for updating network parameters. In lines 2-9, the DRLD-SP agent performs data acquisition to train DNN for each episode. It observes the network state (line 3) and calculates the number of instances required to handle the traffic (line 4). In line 5, the actor network calculates $x_e^s$ using policy defined in Section \ref{Sec:policy}. Then, according to the current policy and state, the action $\mathfrak{a}$ is performed (line 7) to obtain a reward (line 8). A transition of collected information is stored in replay memory $\mathcal{M}$(line 9). Later, the DRLD-SP agent randomly samples a batch of size $\mathcal{N}$ to update the critic network parameters using the loss function (line 10-12). Once the network is trained, the procedure for decision making gets simple and efficient as explained in Algorithm \ref{Alg:DRLD-Decision}. Here, by making use of a trained critic network, the DRLD-SP agent observes the state (Line 1) and performs an action for which quality value is maximum (Line 3). Later, it obtains a reward (Line 4) and observes a new state (Line 5) to facilitate traffic for the next time unit and so on.

\begin{algorithm}
	\DontPrintSemicolon
	\KwInput{Initialize the critic neural network $Q(\omega,\mathfrak{a})$ with parameters $\theta$ and replay memory buffer $\mathcal{M}$}
	\KwInput{Service profile, edge profile}
	
	\For{episode=1,2,3,...., $\mathbb{U}$}
	{	
		\For{t =1 to T}
		{
			Observe the state $\omega$ using (\ref{eq:omega}) \\
			Calculate $\mathcal{I}_s$ for all $s\in S$ using (\ref{eq:instance}) \\
			Calculate $x_e^s$ using actor network (policy function module)\\
			Set $\mathfrak{a}=x_e^s$ \\
			Perform \textit{action} $\mathfrak{a}$ \\
			Obtain reward $\mathcal{R}(\omega,\mathfrak{a})$ using (\ref{eq:reward})\\
			Store transition $[\omega,\mathfrak{a},\mathcal{R}(\omega,\mathfrak{a})]$ in replay memory buffer $\mathcal{M}$ \\
		}
		Sample a batch of $\mathcal{N}$ samples from $\mathcal{M}$ \\
		Set $y_t$ with (\ref{eq:yt}) \\
		Update critic network parameter by minimizing the loss (\ref{eq:loss}) \\
	}
	\KwReturn{The parameters of trained critic DNN}
	\caption{DRLD-SP Network Optimization}
	\label{Alg:DRLD}
\end{algorithm}

\begin{algorithm}
	\DontPrintSemicolon
	\KwInput{Trained critic network with parameters $\theta$}
	\KwInput{Service profile, edge profile}
	Initialize the state $\omega_{\mathfrak{t}}$ \\
	\For{$\mathfrak{t}$=1,2,3,...., }
	{	
		Perform action $\mathfrak{a}=$ arg max $Q(\omega_{\mathfrak{t}}, \mathfrak{a}_{\mathfrak{t}}; \theta)$ \\
		Obtain reward $\mathcal{R}(\omega_{\mathfrak{t}},\mathfrak{a}_{\mathfrak{t}})$\\
		Update $\omega_{\mathfrak{t}} \rightarrow \omega_{\mathfrak{t}+1}$
	}
	\caption{DRLD-SP Decision Making Process}
	\label{Alg:DRLD-Decision}
\end{algorithm}

\section{Performance Evaluation}
\label{sec:results}
In this section, we present performance evaluation results obtained from the extensive simulation of the proposed DRLD-SP algorithm over an IoV network. We start by describing the simulation scenarios for the edge-enabled IoV environment and parameters used to train the optimization model and neural network.  
\subsection{Simulation Setup}
\begin{figure}[htbp]
	\begin{center}
		\includegraphics[width=2.5in,height=1.7in]{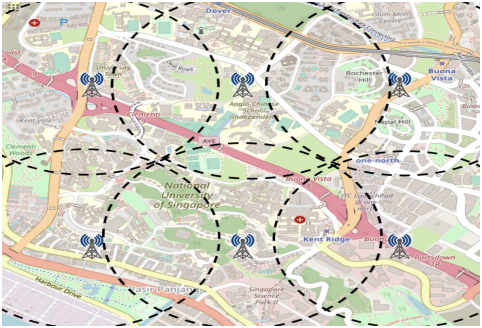}
		\caption{The simulation scenario illustrating quality-coverage of edge nodes}
		\label{fig:SimulationScenario}
	\end{center}
\end{figure}

We use SUMO and MATLAB to set up the simulation environment. SUMO is an open-source simulator, used to simulate a virtual traffic scenario of a realistic vehicular network. In this work, we extract the area of $3km^2$ using Openstreetmaps \cite{OpenStreetMap}. Fig. \ref{fig:SimulationScenario} shows the geographic region and eNBs nodes equipped with MEC servers to provide coverage to the vehicles. The choice of the area is significant as it is present in the center of the city with high traffic densities (Urban environment). Furthermore, the randomTrip application of the SUMO package is used to automatically generate the trips for the vehicles with mobility over the given area of the map. We collect traces of data which helps to generate a 4-tuple service request message dynamically for our algorithm. Table \ref{tab:sim-parameters} lists the parameter values used in the simulation. Different sets of values are chosen for performing multiple experiments. We assume delay critical services and a small threshold is chosen to enforce strict delay constraints. Whereas, the selection of resource unit for $R_s$ and $C_e$ is random. Experiments are performed for different sets (by choosing the lowest values as well as the highest values) of $R_s$ and similar performance trends are observed. \par
The implementation of the DRLD-SP agent is carried out using MATLAB. For the neural network design, we conducted a comprehensive experimental study to find the best hyperparameters. We use 3-layer fully-connected feedforward critic network. The size of the input layer is the same as the dimension of the network input states. It has 3 hidden layers, each with 256, 64, and 32 neurons respectively. We use hyperbolic tangent sigmoid for activation of hidden layers. The output layer is a single neuron that expresses the Q-value. We use the linear transfer function for the activation of the output layer. To avoid overfitting, the learning rate of 0.01 is used to train a network. We set up the size of a batch as 100. The maximum number of episodes performed to train a network is 5000 with each episode having a maximum of 20 iterations. The parameters of the critic network are updated every 5 time slots. It achieves accuracy of 90\%+ in 4.6min. All experiments are evaluated on a system with Intel Corei5 2GHz and 8GB RAM.
\begin{table}[htbp]
	\centering
	\caption{Simulation Parameters}
	\begin{tabular}{ll}
		\toprule
		\textbf{Parameters} & \textbf{Value} \\
		\midrule
		$S$     & 8 \\
		$V$     & 200 \\
		$E$    & 6 \\
		$R_s (Unit)$ & [10 15 20 25 30 35 40 45] \\
		$C_e (Unit)$ & [60 70 80 90 100 100] \\
		$D_s (ms)$ & [10 10 10 10 12 12 12 12] \\
		$\alpha$ & [0.2 0.4 0.6 0.8 1] \\		
		$\mathbb{C}$ & 15 \\
		$\mathcal{N}$ & 100 \\
		$\mathfrak{t}$  & 1 to 600 \\
		\bottomrule
	\end{tabular}%
	\label{tab:sim-parameters}%
\end{table}%

\subsection{Performance Metrics}
\begin{figure*}[hbt!]
	\centering
	\begin{subfigure}{.19\textwidth}
		\centering
		\includegraphics[width=1.5in]{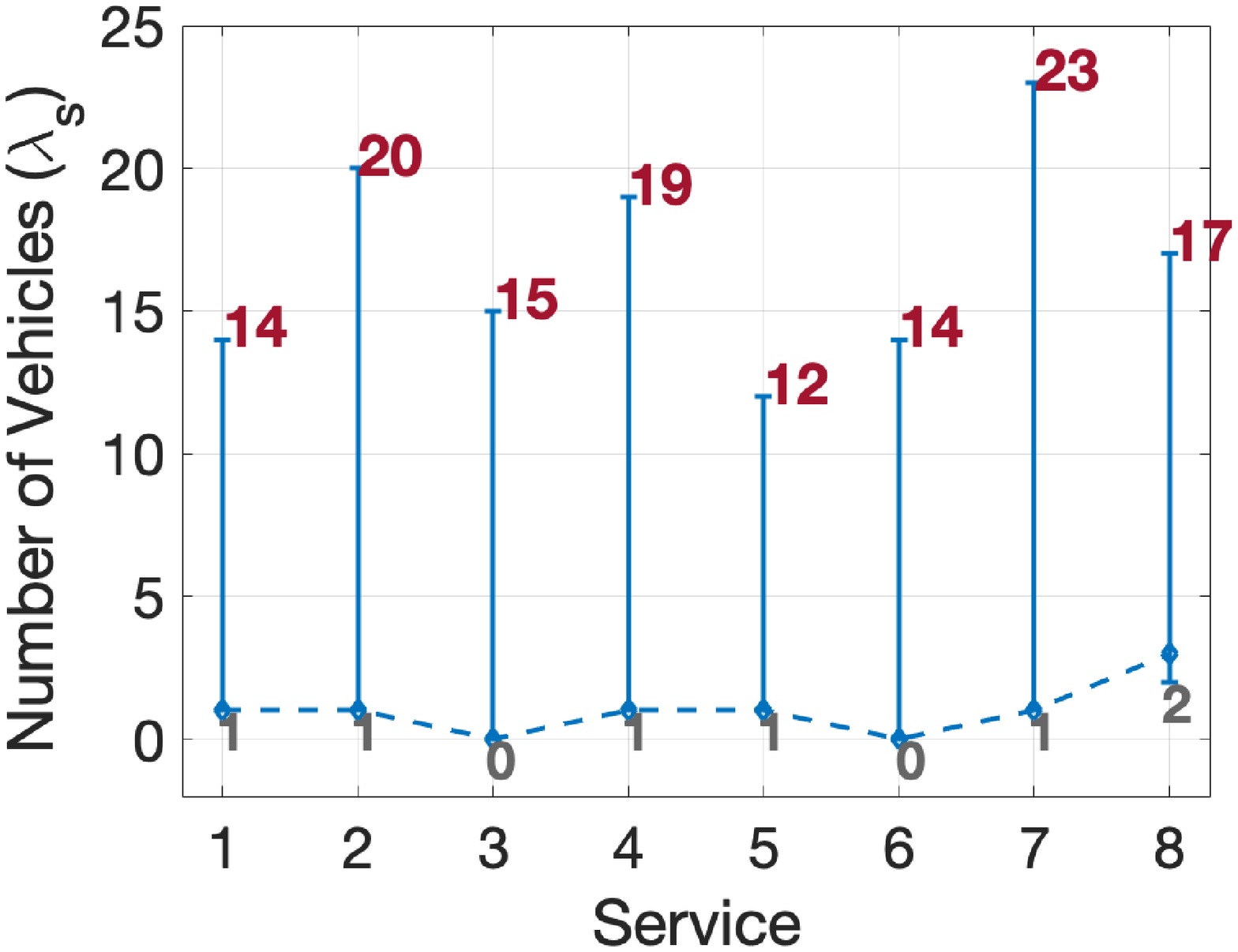}  
		\caption{Trial 01}
		\label{fig:data1}
	\end{subfigure}
	\begin{subfigure}{.19\textwidth}
		\centering
		\includegraphics[width=1.5in]{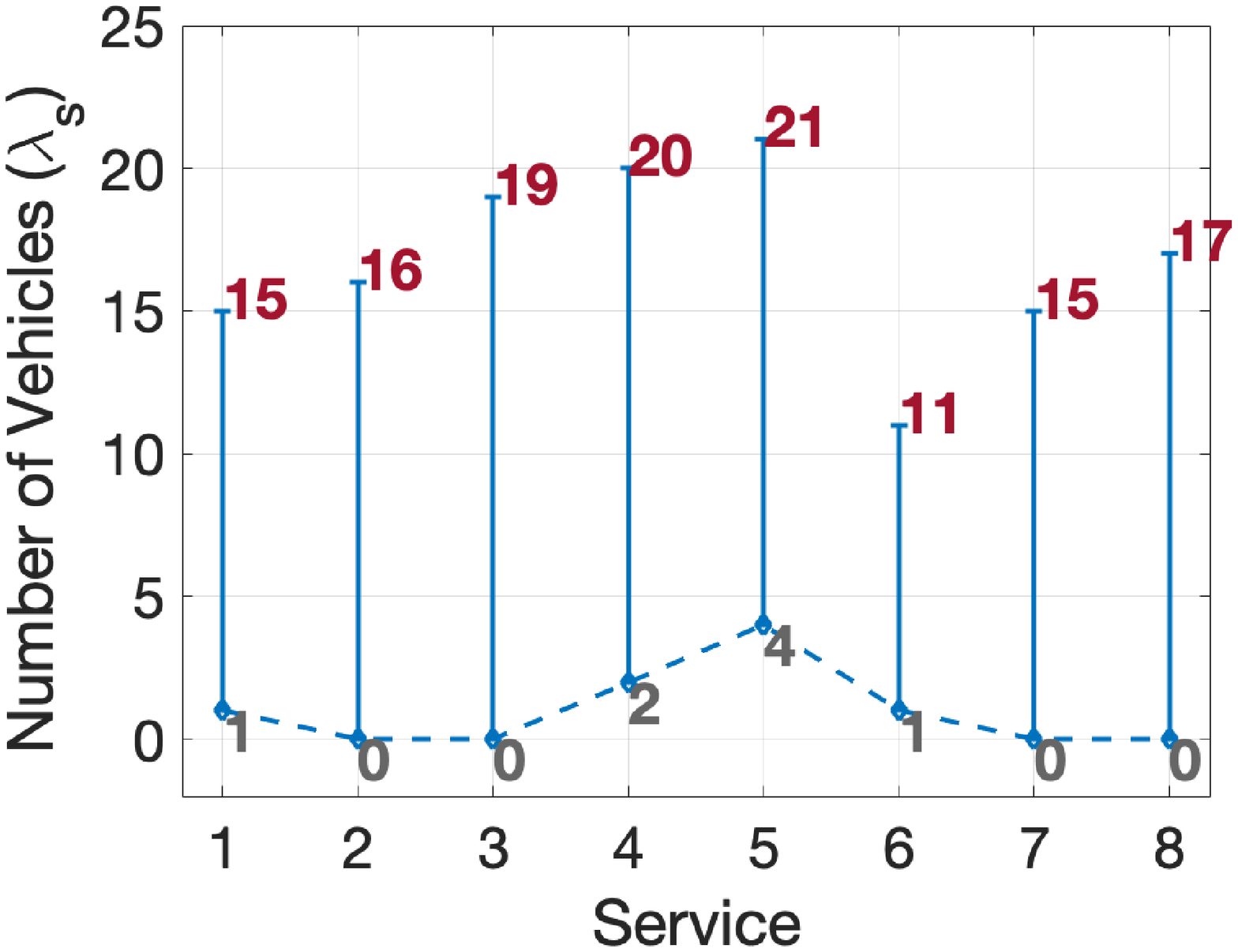}  
		\caption{Trial 02}
		\label{fig:data2}
	\end{subfigure} 
	\begin{subfigure}{.19\textwidth}
		\centering
		\includegraphics[width=1.5in]{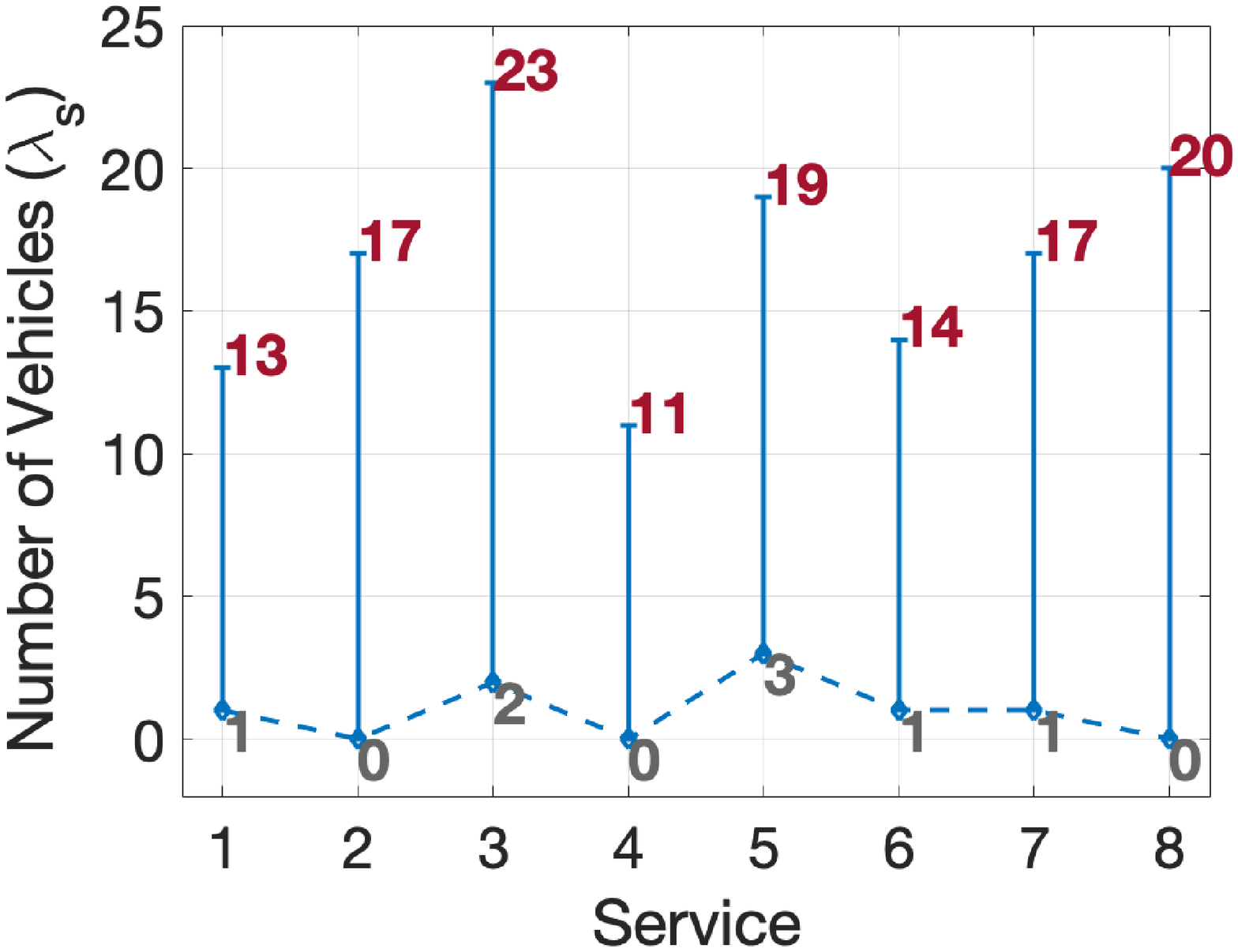}  
		\caption{Trial 03}
		\label{fig:data3}
	\end{subfigure}
	\begin{subfigure}{.19\textwidth}
		\centering
		\includegraphics[width=1.5in]{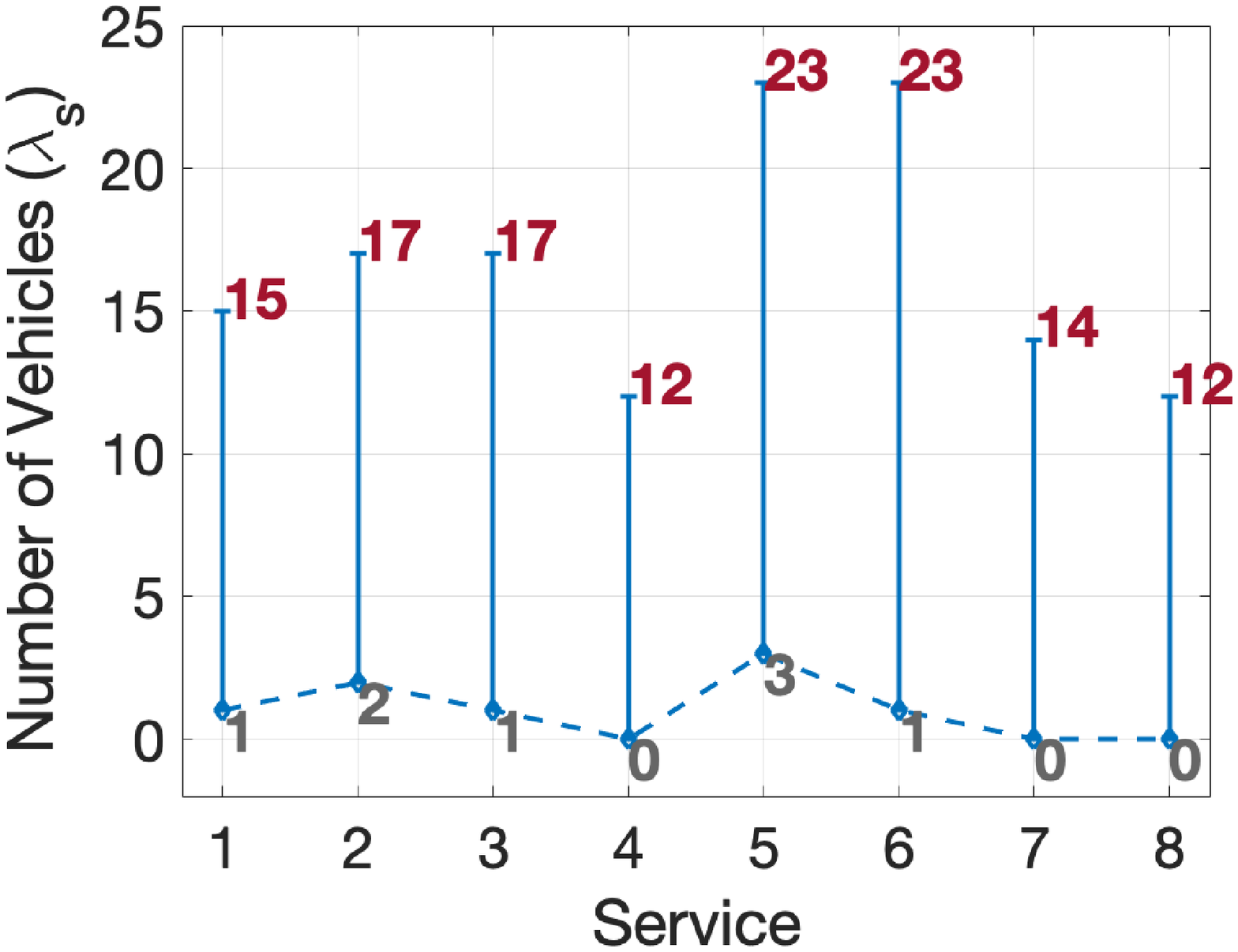}  
		\caption{Trial 04}
		\label{fig:data4}
	\end{subfigure} 
	\begin{subfigure}{.19\textwidth}
		\centering
		\includegraphics[width=1.5in]{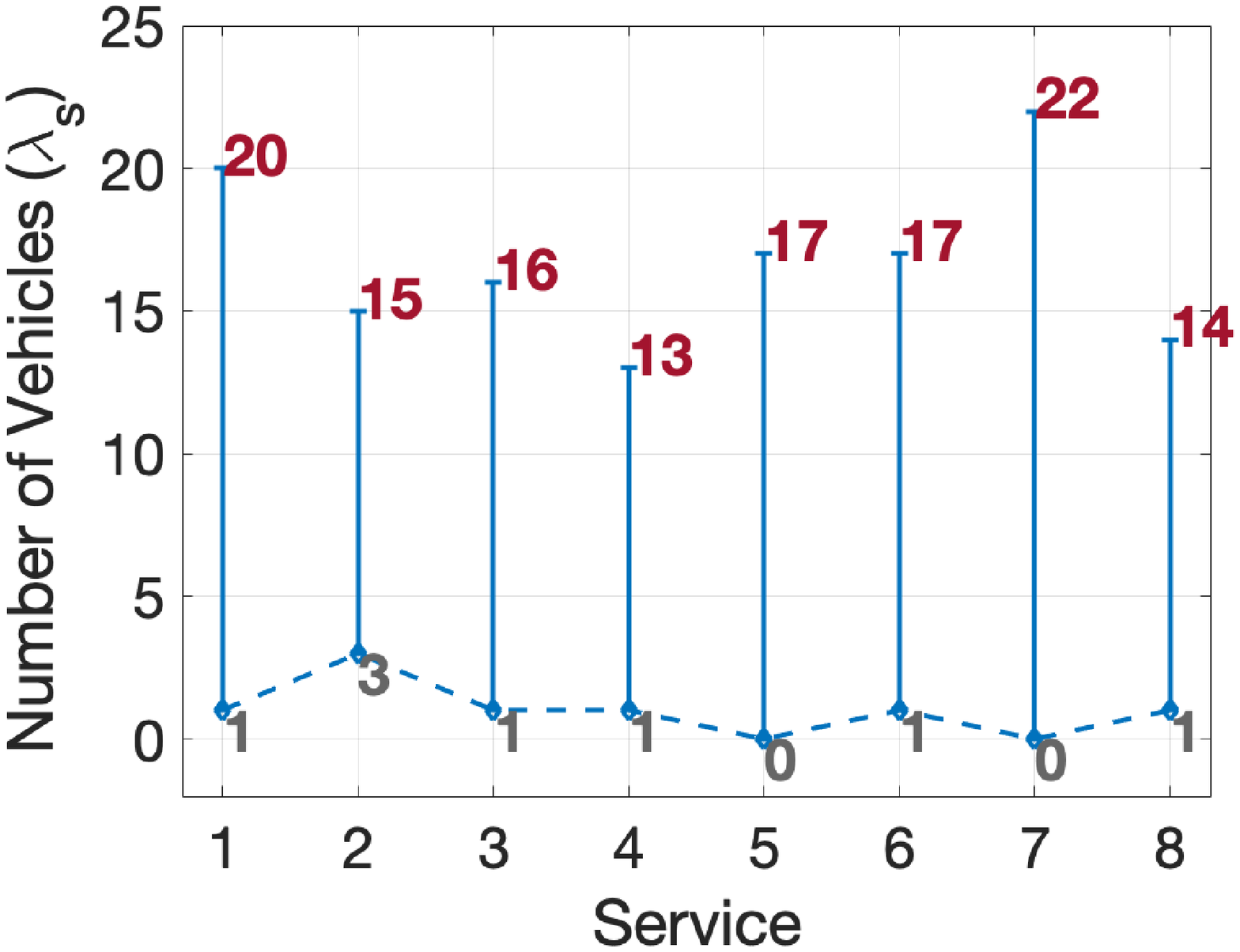}  
		\caption{Trial 05}
		\label{fig:data5}
	\end{subfigure}
	\caption{Number of vehicles requesting for a service at a time (min-to-max bar)}
	\label{fig:dataset}
\end{figure*}
\begin{figure*}[htbp]
	\centering
	\begin{subfigure}{.19\textwidth}
		\centering
		\includegraphics[width=1.5in,height=1.18in]{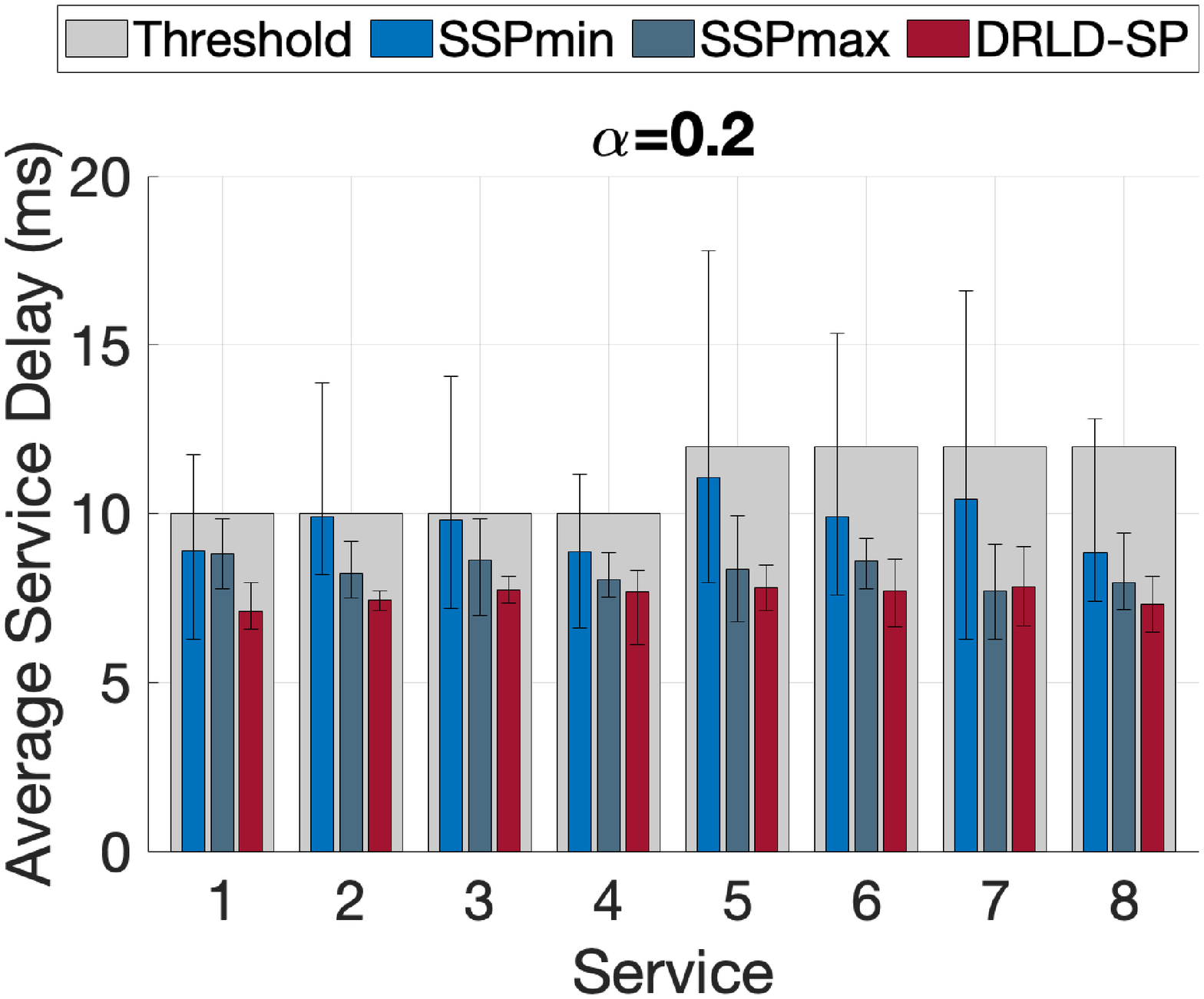}  
		\caption{$\alpha=0.2$}
		\label{fig:DEB02}
	\end{subfigure}
	\begin{subfigure}{.19\textwidth}
		\centering
		\includegraphics[width=1.5in,height=1.18in]{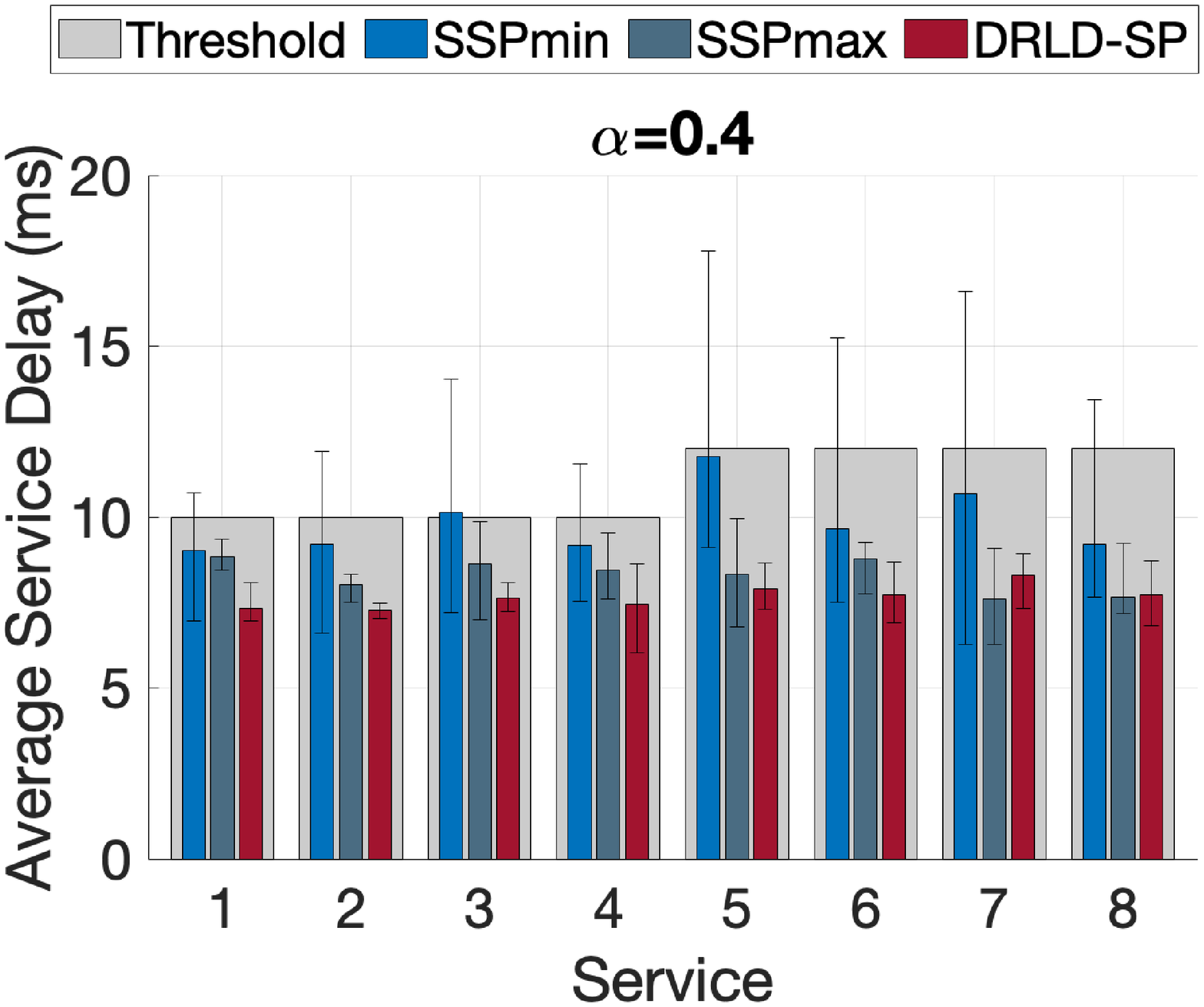}  
		\caption{$\alpha=0.4$}
		\label{fig:DEB04}
	\end{subfigure}
	\begin{subfigure}{.19\textwidth}
		\centering
		\includegraphics[width=1.5in,height=1.18in]{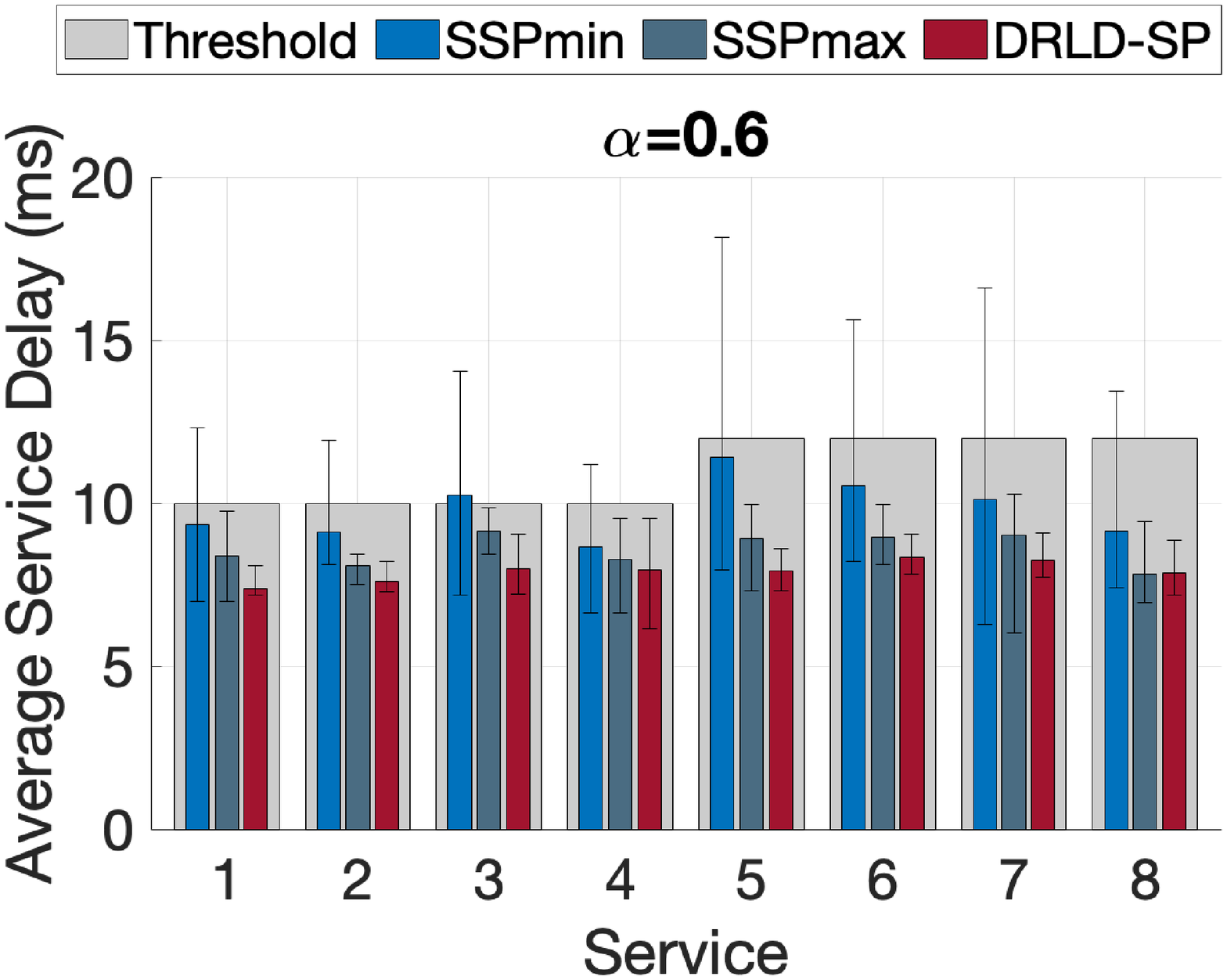}  
		\caption{$\alpha=0.6$}
		\label{fig:DEB06}
	\end{subfigure}
	\begin{subfigure}{.19\textwidth}
		\centering
		\includegraphics[width=1.5in,height=1.18in]{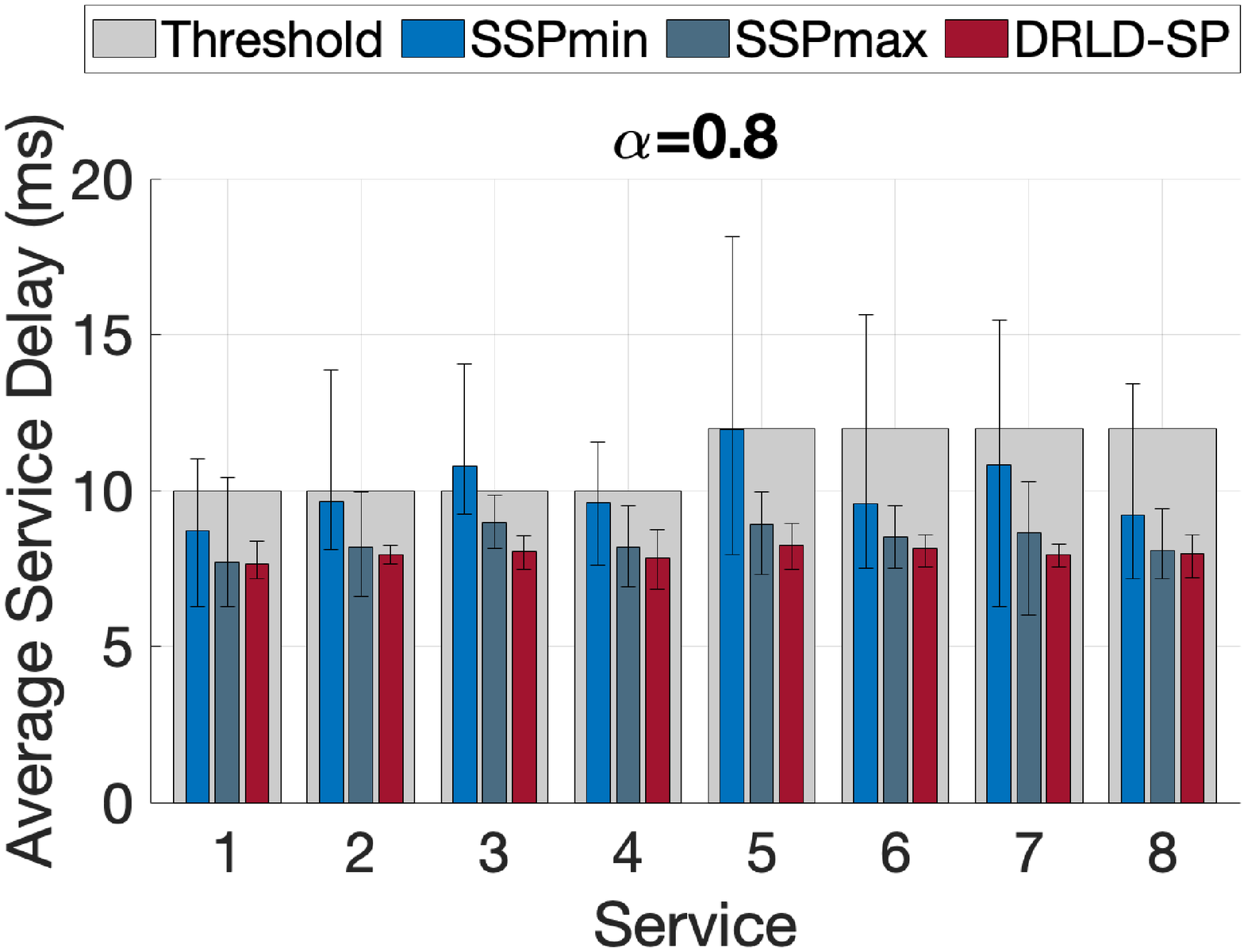}  
		\caption{$\alpha=0.8$}
		\label{fig:DEB08}
	\end{subfigure}
	\begin{subfigure}{.19\textwidth}
		\centering
		\includegraphics[width=1.5in,height=1.18in]{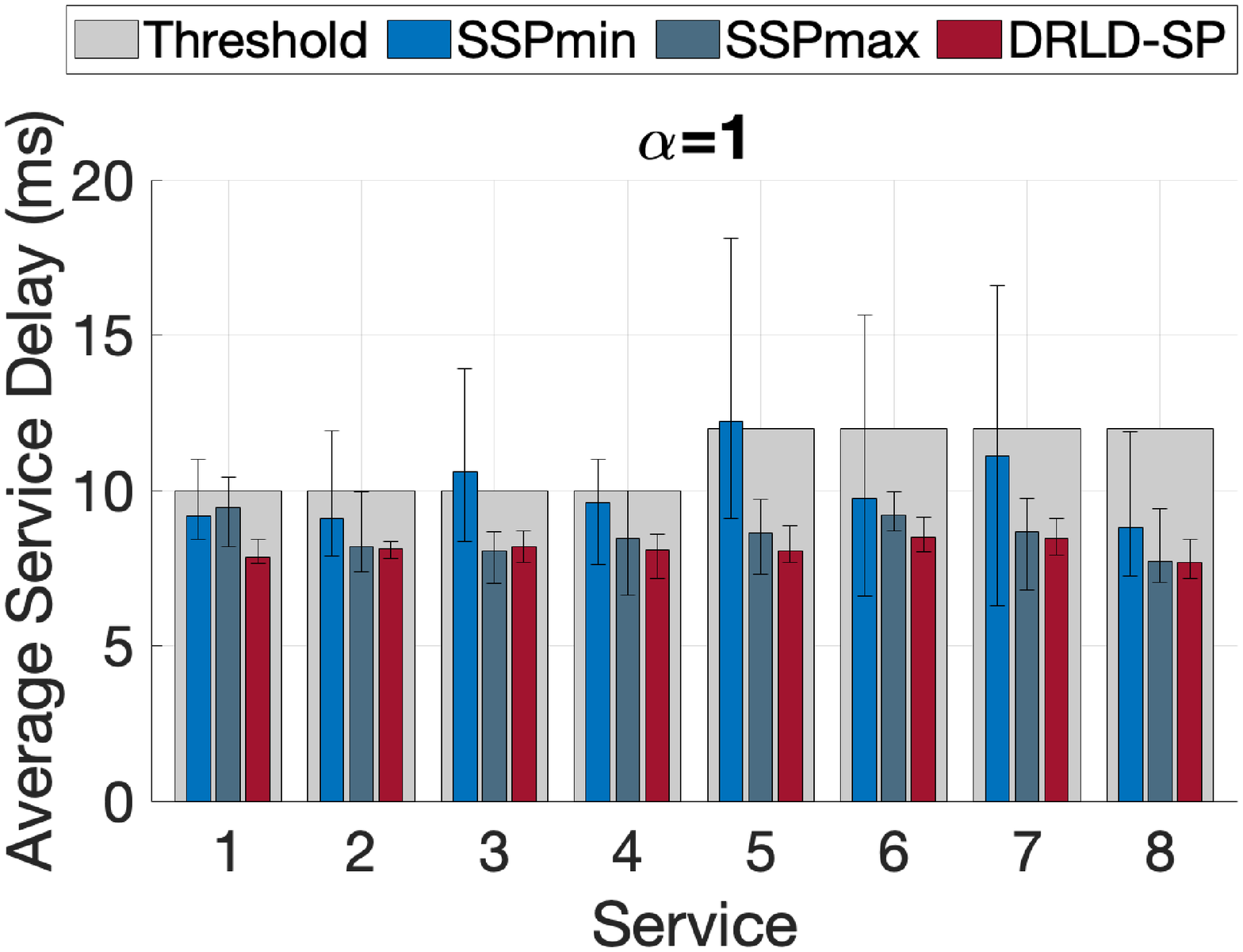}  
		\caption{$\alpha=1$}
		\label{fig:DEB10}
	\end{subfigure}
	\caption{Average service delay}
	\label{fig:delayEB}
\end{figure*}

To verify the performance of our proposed DRLD-SP mechanism, we use the following metrics. 

\begin{itemize}
	\item \textit{Average Service Delay:} It is the average delay experienced for different services by the vehicles.
	\item \textit{Edge Resource Usage:} It is the ratio between the resources that $\mathcal{I}_s$ instances of service $s$ will consume after placement and the available resources at the edge node. This metric focuses on the minimum usage of limited edge resources.
	\item \textit{Fairness:} The fairness of server utilization is a representation of fair and efficient resource consumption. It also determines the level of load balancing among different edge nodes. We use Jain's index as a fairness measure in this work \cite{jain_index}. The edge server utilization is fairer when Jain$'$s index is closer to 1. 
	\item \textit{Service Instance Utilization:} We define a utility function for service instance utilization as:
	\begin{equation}
	\overline{\mathcal{U}_s}=\frac{1}{T}\sum_{t=0}^{T}\left(\frac{\lambda_s(t)}{\mathcal{I}_s\mathbb{C}} \right)
	\end{equation}
	Where $\mathcal{I}_s$ is the total number of instances deployed for service $s$. We use this metric to show the efficiency of placement algorithm in terms of utilizing the deployed service instances at the edge. 
	\item \textit{Service Satisfaction:} We define a utility function for service satisfaction as:
	\begin{equation}
	\zeta_s(t)=
	\begin{cases}
	1 & \lambda_s\le \mathbb{C}  \\
	\frac{\mathbb{C}}{\lambda_s} & \lambda_s> \mathbb{C}
	\end{cases}
	\end{equation}
	where $\lambda_s$ is the number of vehicles requesting for service $s$ at time $t$ and $\mathbb{C}$ is the number of vehicles handled by an instance of service per unit time. The service satisfaction shows how efficient service placement decision is, and helps to be aware of the proportion of traffic able to get service from the edge without service congestion and waiting delay. 
	\item \textit{Average Instances Installed:} This metric represents the average of instances deployed for each service. It helps to measure the performance of the placement algorithm in terms of the average of instances installed throughout the period while facilitating the IoV traffic. 
	\item \textit{Re-placement Cost:} This metric gives the number of times the algorithm re-optimizes within the total duration of the experiment. The higher value of re-placement cost (or migration cost) means more optimizations and re-placements implying more service interruptions and network performance degradation.   
\end{itemize}

\subsection{Baseline Algorithms for Comparison}
\label{Sec:compAlg}
We evaluate the performance of our proposed dynamic service placement DRLD-SP algorithm against existing one-time placement static algorithm \cite{Edge-enabledV2X} and mobility-aware dynamic schemes \cite{D1latencyaware,D3cloud}. We suitably modify these schemes (keeping the underlying approach unaffected) to suit our problem context for a fair comparison with our algorithm. \par 
A static solution is a baseline technique that fixes servers for hosting services by performing a one-time ILP placement solution. Earlier works that provide ILP-based static solutions for service placement and edge resource allocation were briefly discussed in Section \ref{sec:relatedwork}. As a baseline, we use the static placement scheme developed in \cite{Edge-enabledV2X}. To evaluate the efficiency of handling the varying demand for services from vehicles we compare our algorithm with two versions of static placement; one is a static placement where it deploys 1 instance for each service, and another is to deploy 2 instances for each service to handle the maximum traffic. We plot the simulation input data in Fig. \ref{fig:dataset} to show variation in the request for services by vehicles (min to max). Considering this, the number of instances required to handle the maximum load is 2. Since 1 instance can handle 15 vehicles at a time (i.e. $\mathbb{C}=15$, shown in Table \ref{tab:sim-parameters}), 2 instances can handle a maximum of 30 vehicles at a time. We formally call two different versions of static service placement as, $\boldsymbol{SSP_{min}}$ (i.e. $\mathcal{I}_s$=1 for all services), and $\boldsymbol{SSP_{max}}$ (i.e. $\mathcal{I}_s$=2 for all services). Here, SSP stands for static service placement. \par 
\begin{figure}[htbp]
	\centering
	\includegraphics[width=2.8in,height=2in]{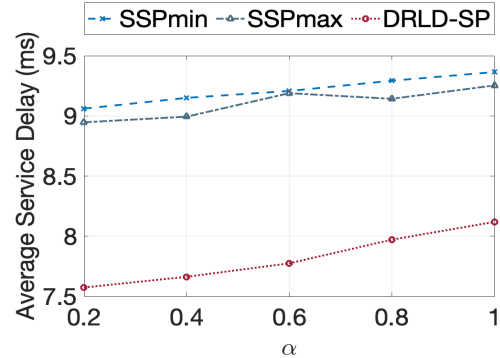}  
	\caption{Average service delay vs. $\alpha$}
	\label{fig:delayvsAlpha}
\end{figure}
We noted that our proposed \textbf{\textit{DRLD-SP}} algorithm considers the traffic mobility along with service dynamics while making a decision on the number of instances for a service. Therefore, we also compare our DRLD-SP with existing dynamic schemes which are termed as always-reoptimize \textbf{\textit{(AR)}} \cite{D1latencyaware} and threshold-based reoptimization \textbf{\textit{(TBR)}} \cite{D3cloud} in this work. The performance of \textbf{AR} and \textbf{TBR} is also compared in terms of min (i.e. $\mathcal{I}_s$=1) and max (i.e. $\mathcal{I}_s$=2) service instance case. We call them accordingly as $\boldsymbol{AR_{min}}$, $\boldsymbol{AR_{max}}$, $\boldsymbol{TBR_{min}}$ and $\boldsymbol{TBR_{max}}$. These works were briefly discussed in Section \ref{sec:relatedwork}. For TBR comparison, we use a threshold of 9 to satisfy and keep the delay below the minimum threshold ($D_s$) requirement.\par 
\begin{figure*}[htbp]
	\centering
	\begin{subfigure}{.19\textwidth}
		\centering
		\includegraphics[width=1.5in,height=1.18in]{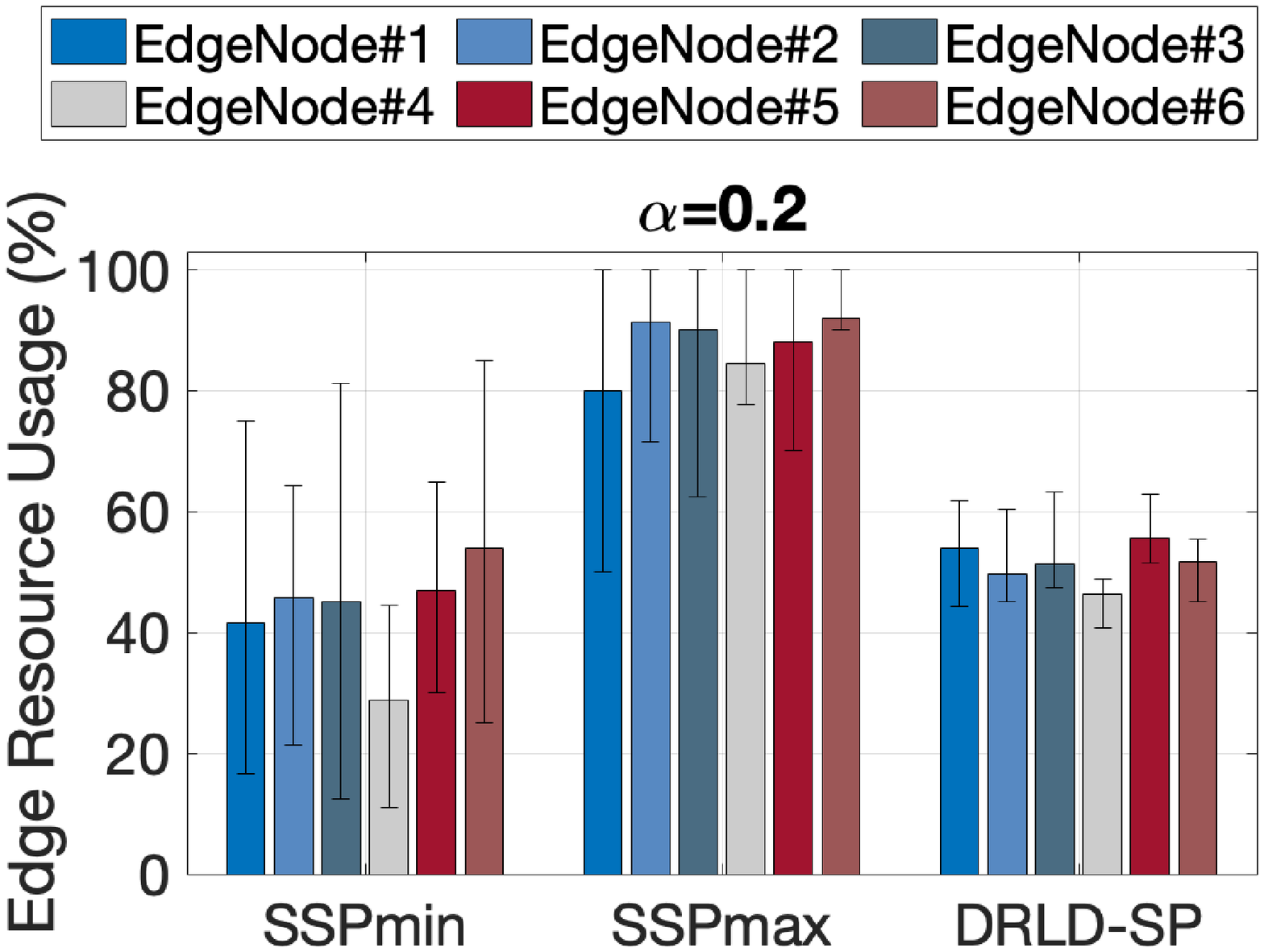}  
		\caption{$\alpha=0.2$}
		\label{fig:REB02}
	\end{subfigure}
	\begin{subfigure}{.19\textwidth}
		\centering
		\includegraphics[width=1.5in,height=1.18in]{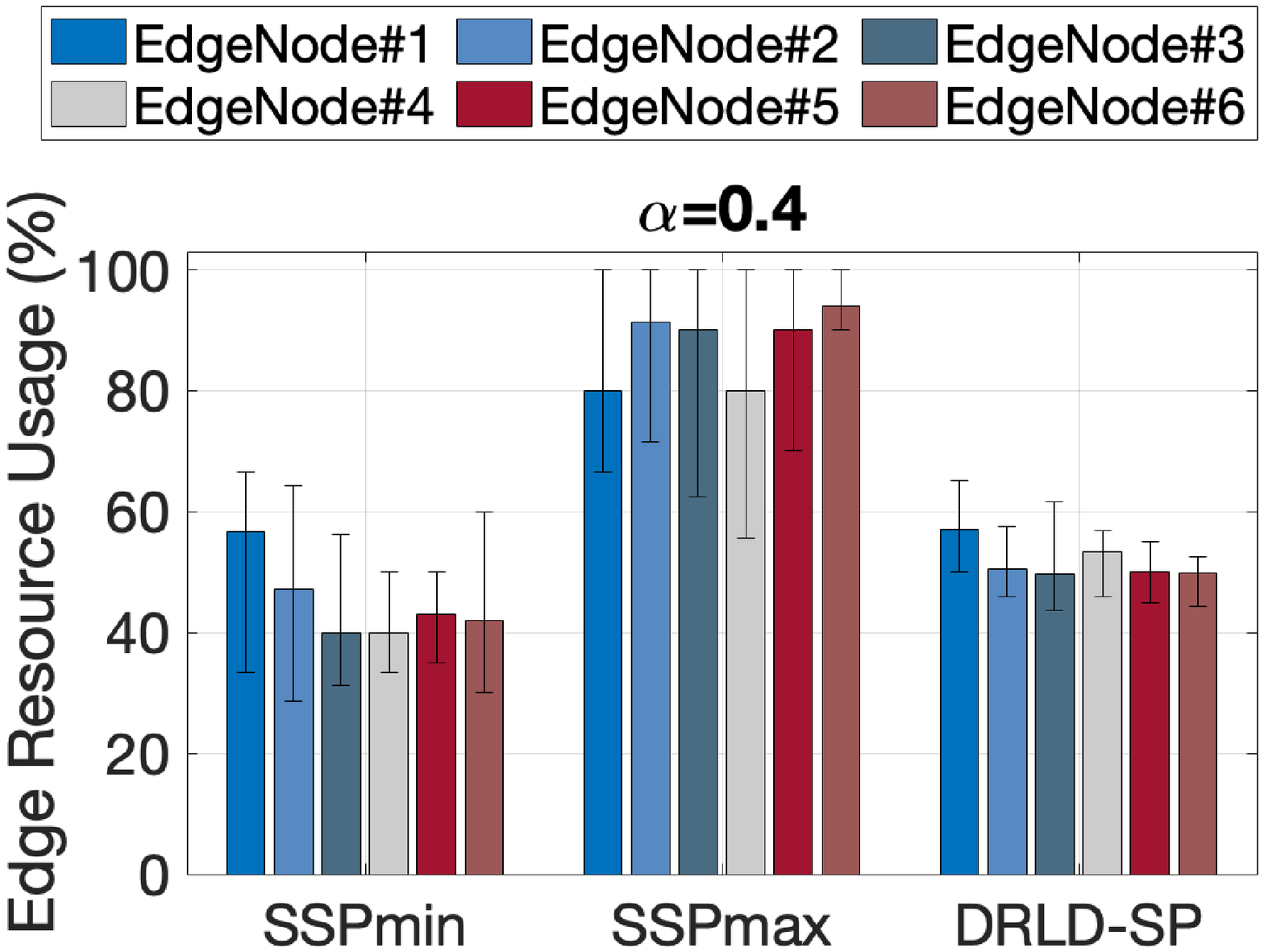}  
		\caption{$\alpha=0.4$}
		\label{fig:REB04}
	\end{subfigure}
	\begin{subfigure}{.19\textwidth}
		\centering
		\includegraphics[width=1.5in,height=1.18in]{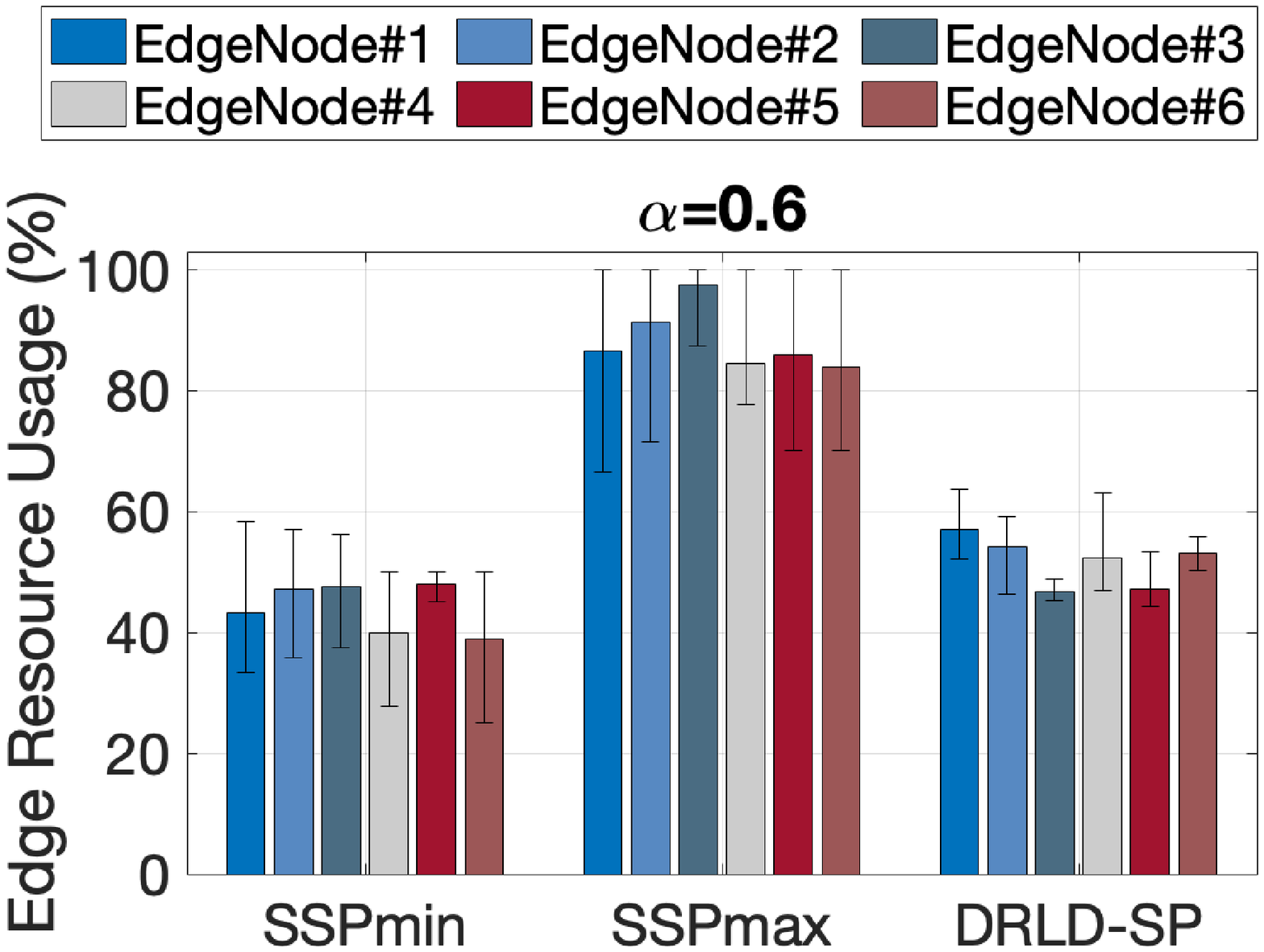}  
		\caption{$\alpha=0.6$}
		\label{fig:REB06}
	\end{subfigure}
	\begin{subfigure}{.19\textwidth}
		\centering
		\includegraphics[width=1.5in,height=1.18in]{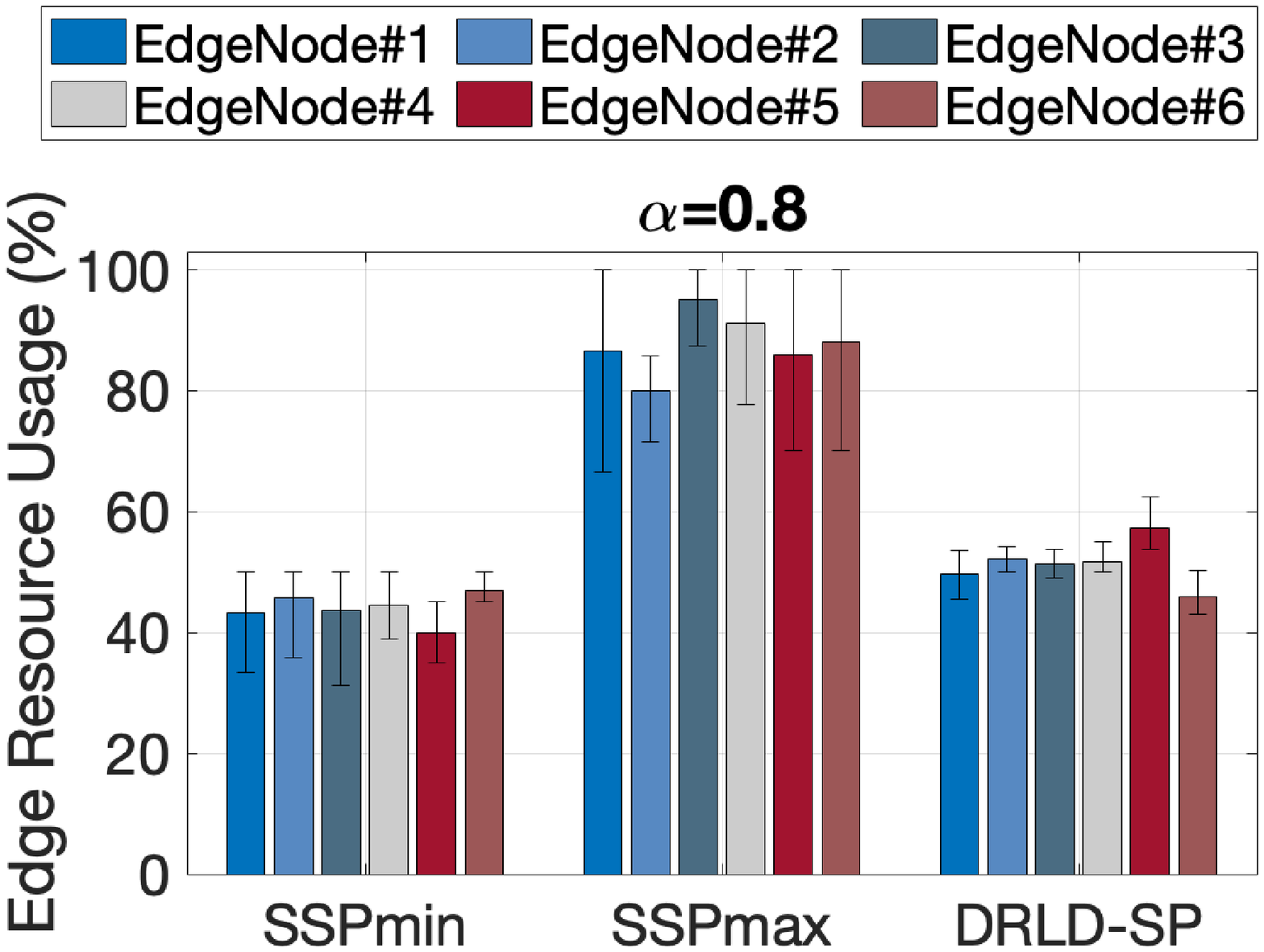}  
		\caption{$\alpha=0.8$}
		\label{fig:REB08}
	\end{subfigure}
	\begin{subfigure}{.19\textwidth}
		\centering
		\includegraphics[width=1.5in,height=1.18in]{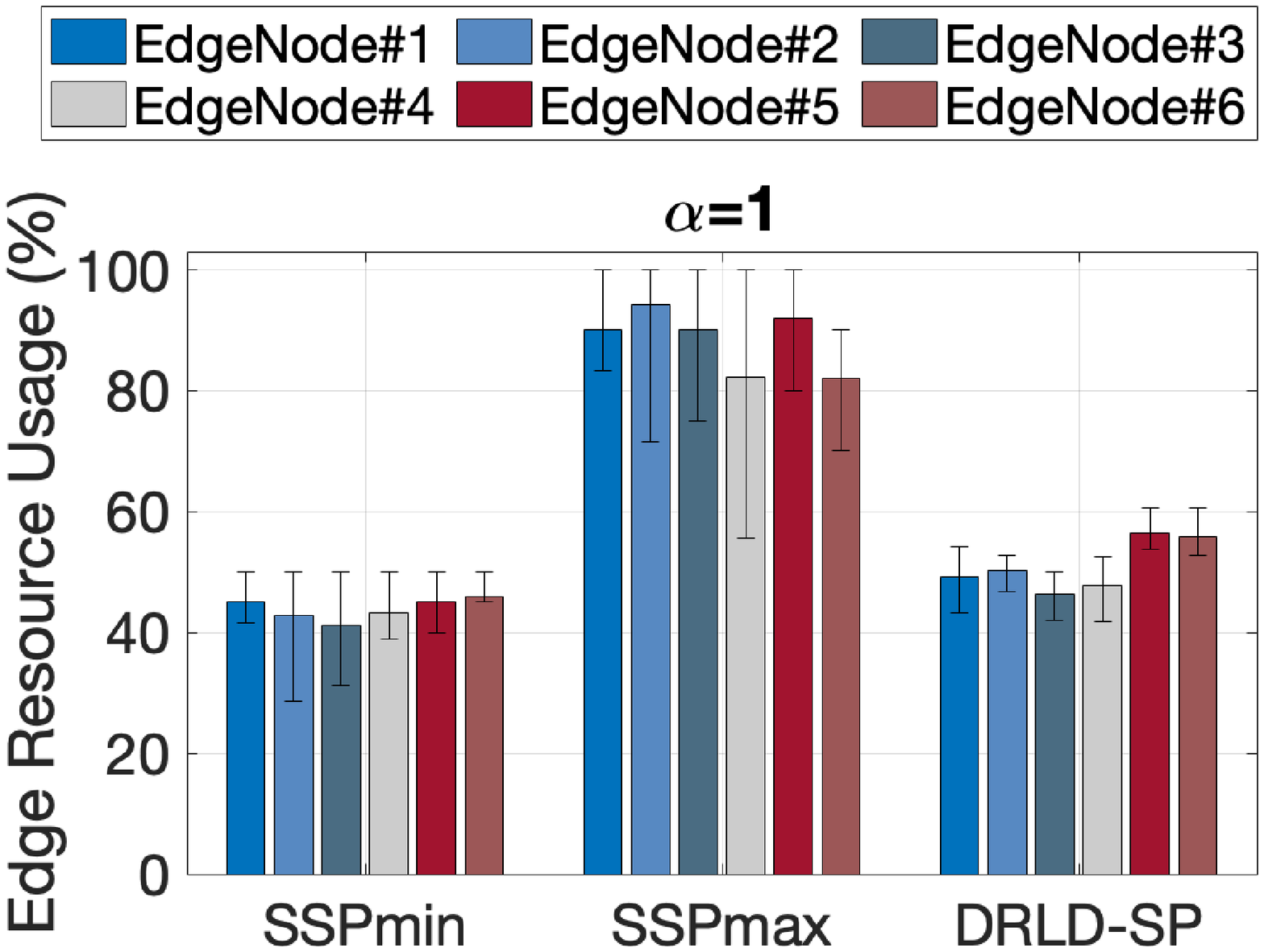}  
		\caption{$\alpha=1$}
		\label{fig:REB10}
	\end{subfigure}
	\caption{Edge resource usage}
	\label{fig:compR}
\end{figure*}
We carry out experiments (trials) five times with different random seeds and for each trial, we vary $\alpha$ from 0.2 to 1 (as shown in Table \ref{tab:sim-parameters}) to study the relative importance of minimizing the maximum of resource usage versus service delay. We present the average of five trials. We use error bars within plots to show minimum-to-maximum variations observed around the average value by performing five trials.

\subsection{Results}
\subsubsection{Performance of proposed DRLD-SP framework}
In this section, we briefly discuss the performance of our proposed DRLD-SP framework against two versions of SSP using different evaluation metrics. In the first place, Fig. \ref{fig:delayEB} illustrates the average delay experienced for different services by the vehicles. We compare the average service delay of our proposed DRLD-SP framework against two versions of static service placement ($SSP_{min}$ and $SSP_{max}$), as discussed in Section \ref{Sec:compAlg}. It can be observed that most of the time the average delay observed by vehicles for $SSP_{min}$ is the highest of all. This is because a limited portion of service requests can be handled by one instance of service, as more requests arrive, the queuing delay increases. However, the $SSP_{max}$ has the ability to handle the maximum load but the delay observed in $SSP_{max}$ is also greater than our proposed DRLD-SP framework. This is due to the fact that the vehicular environment is not stationary. The high mobility of nodes and constantly changing topology requires continual learning of the environment (as in DRLD-SP) to provide a better average delay for each service. Another observation from Fig. \ref{fig:delayEB} is that both SSP deployments are not able to satisfy the delay threshold requirement for some services. In contrast, the delay for DRLD-SP is always well below its threshold for all trials. \par 
We evaluate the performance in terms of average delay for all services and all trails against $\alpha$ (i.e. relative importance of edge resource vs delay), in Fig. \ref{fig:delayvsAlpha}. Higher the value of $\alpha$, the lesser the importance of delay. It can be observed that the trend for delay vs. $\alpha$ is linear in DRLD-SP, from the fact of decreasing importance for the delay. However, in SSP for minimum as well as maximum load handling capability, the trend is erratic. The delay for $\alpha=0.2$ is lesser, compared to $\alpha=1$, but it is not linear. This will make it difficult for the service provider on deciding hyper parameters of SSP deployment because of unpredictable performance. With this, we show the effectiveness of adopting a learning method by the dynamic approach.
\par
Fig. \ref{fig:compR} shows the average edge resources consumed by three different types of placement methods. In SPP, to accommodate max load (i.e. $SSP_{max}$), the usage of resources is very high. On the contrary, the $SSP_{min}$ consumes resources lower to a small degree than DRLD-SP but adds-up waiting delays and service request congestions. Our proposed DRLD-SP intends to utilize edge-server resources more effectively accommodating the same demand (as carried out by $SSP_{max}$), but with low usage of edge resources. \par
\begin{figure}[htbp]
	\centering
	\includegraphics[width=2.8in,height=2in]{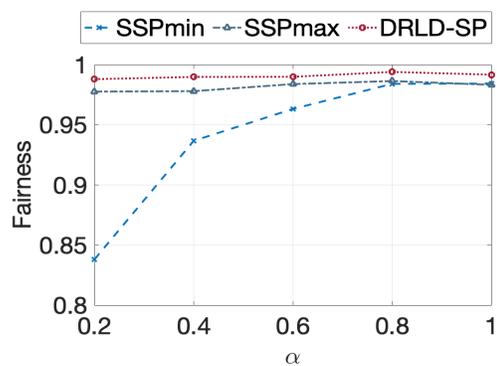}  
	\caption{Fairness}
	\label{fig:fairness}
\end{figure}

 
Further, we compare the balanced spread of service resources against changing values of $\alpha$. We plot average fairness calculated from five trials in Fig. \ref{fig:fairness} to represent the load balance among the edge nodes. The balanced spread of service resources among edge servers should increase with the increasing $\alpha$ and help to prevent the saturation/congestion at any single server given the limited resources at the servers. With SPP, the $SSP_{min}$ is exhibiting better performance only for higher values of $\alpha$. In $SSP_{max}$, the fairness is slightly lower than DRLD-SP but considering the fact that the resources of all servers in $SSP_{max}$ are always highly-consumed so its difficult to evaluate the performance of fairness. In the case of our proposed DRLD-SP, the spread of service resources exhibits substantially higher fairness for all values of $\alpha$, and mitigates the load imbalance and resource wastage problem across the edge nodes, while satisfying the delay requirements. As a matter of fact, the inefficient usage of resources not only results in wastage but also forces future service demands to be accessed from the network core that will incur higher delay leading to lower performance. This demonstrates the effectiveness of the proposed DRLD-SP in edge resource usage and service delay for the proposed IoV network with limited edge resources.

\par 
\begin{figure}[htbp]
	\centering
	\includegraphics[width=2.8in,height=2.25in]{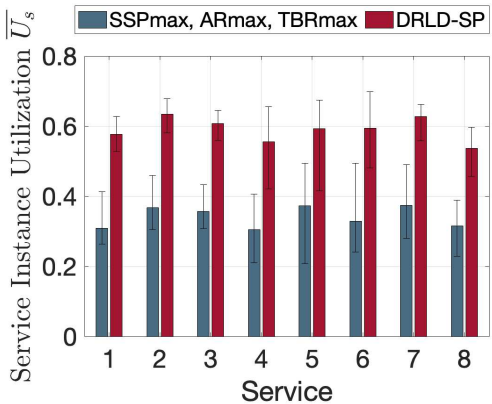}  
	\caption{Service instance utilization}
	\label{fig:SInstance}
\end{figure}
\begin{figure}[htbp]
	\centering
	\includegraphics[width=2.8in,height=2in]{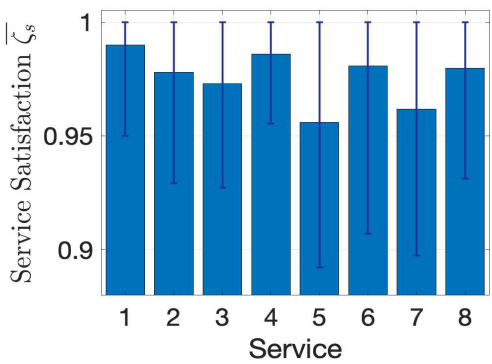}  
	\caption{Service satisfaction for $SSP_{min}$, $AR_{min}$ and $TBR_{min}$}
	\label{fig:ServiceSatisfaction}
\end{figure}
We further study the performance of DRLD-SP in terms of service instance utilization and service satisfaction against maximal and minimal resource-consuming frameworks, respectively. We plot service instance utilization comparing our method with $SSP_{max}$, $AR_{max}$ and $TBR_{max}$ in Fig. \ref{fig:SInstance}. The performance for $SSP_{max}$, $AR_{max}$ and $TBR_{max}$ is similar because all the maximum-utilization scenarios use two service instances for each type of service. The resources available at the edge are limited and significant for latency-sensitive IoV networks. Once the resources are used while placing a service, it's important to utilize it the utmost to avoid any wastage of resources. As depicted in Fig. \ref{fig:SInstance}, the average service instance utilization by $SSP_{max}$, $AR_{max}$ and $TBR_{max}$ is low. On the other hand, our proposed DRLD-SP utilizes service instances more efficiently. \par
\begin{figure}[htbp]
	\centering
	\includegraphics[width=2.8in,height=2.2in]{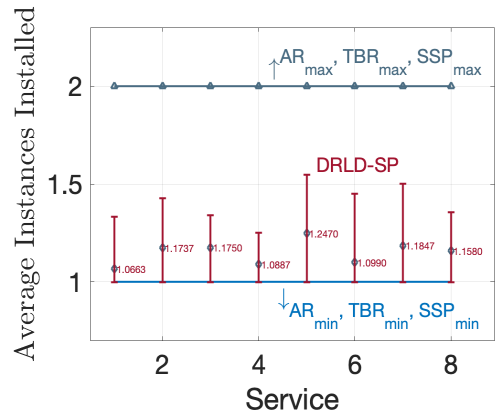}  
	\caption{Average instances installed}
	\label{fig:AvgInstance}
\end{figure}
On the contrary, if minimum resource usage is considered, the wastage of resources can be minimized but it has the drawback of low service satisfaction. Fig. \ref{fig:ServiceSatisfaction} depicts the results for service satisfaction for $SSP_{min}$, $AR_{min}$, and $TBR_{min}$. The service satisfaction is degraded when demand increases for any particular service. We plot the average for each service in Fig. \ref{fig:ServiceSatisfaction} and the error bar shows max-to-min variation in service satisfaction value for five experimental trials. This is because a smaller portion of service requests can be addressed by one instance of service. In the case of DRLD-SP, the service satisfaction is always 1 because of its dynamic nature where the placement decision accommodates the varying demand of vehicles. The results imply that DRLD-SP uses a good policy to place services over the edge without giving any downsides to the user or service provider. \par 
In Fig. \ref{fig:AvgInstance}, we compare the average instances installed by different types of service placement methods. As can be observed from the figure, our proposed DRLD-SP placement intends to utilize edge resources more effectively accommodating the same demand (as carried out by $SSP_{max}$, $AR_{max}$ and $TBR_{max}$), but with lower number of instances. Moreover, for $SSP_{min}$, $AR_{min}$, and $TBR_{min}$, the number of instances for all services is always 1 but it leads to congestions and queuing delays and hence unable to fulfill delay threshold requirements. \par 
\subsubsection{Impact of abrupt changes to the environment}
\begin{figure}[htbp]
	\centering
	\includegraphics[width=2.8in,height=2.1in]{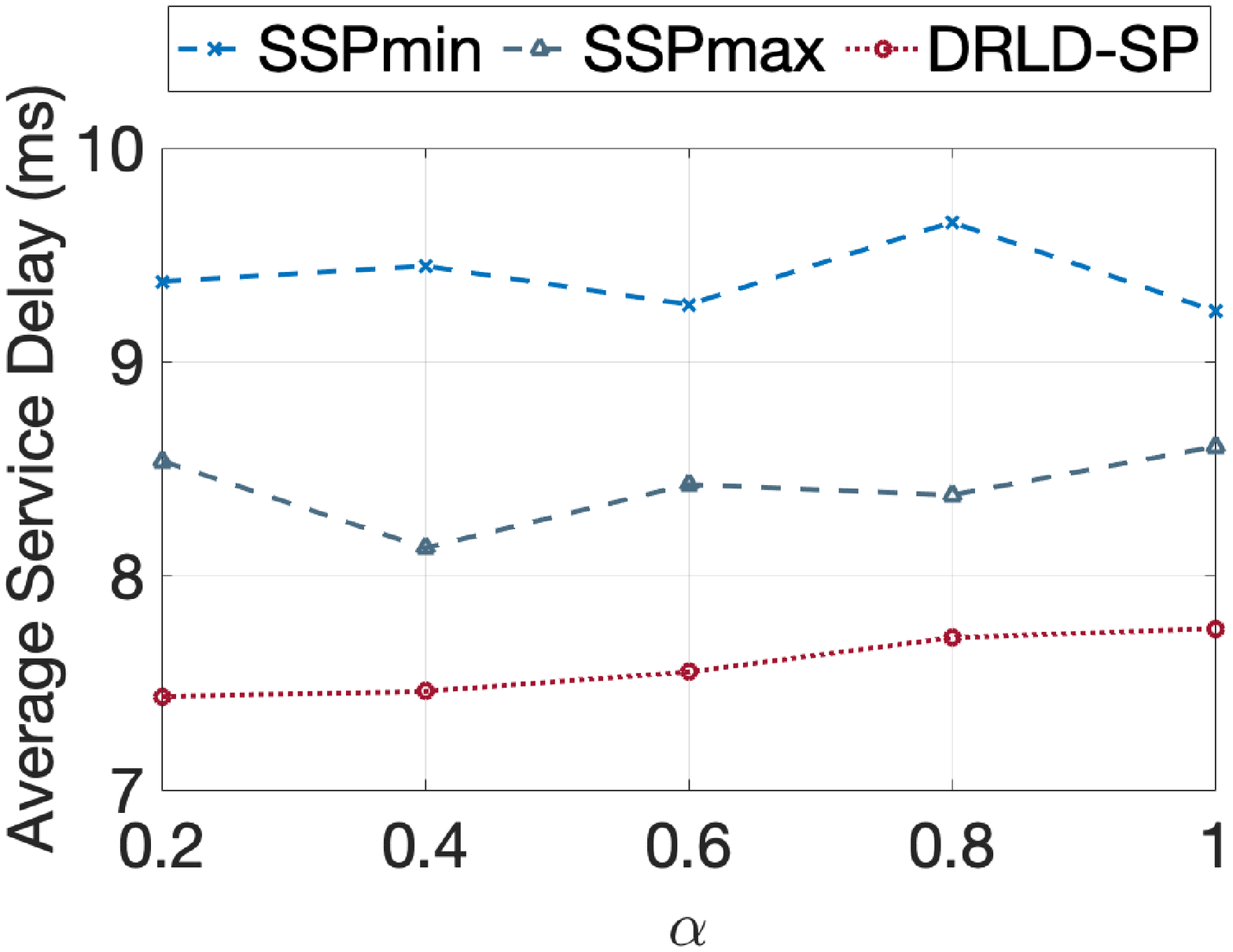}  
	\caption{Average service delay}
	\label{fig:abruptdelay}
\end{figure}

\begin{figure}[htbp]
	\centering
	\includegraphics[width=2.8in,height=2.1in]{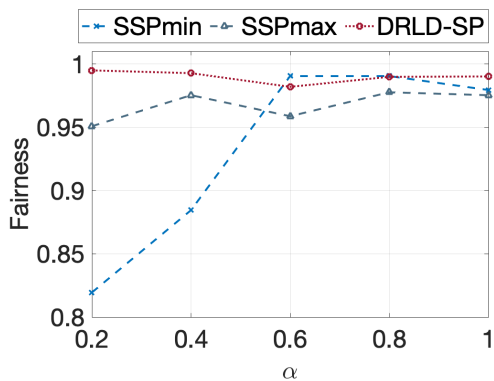}  
	\caption{Fairness}
	\label{fig:abruptfairness}
\end{figure}

In the previous set of experiments, we considered a smooth traffic scenario in which vehicular density changes gradually with the increase in the number of vehicles from 1 to 200\textsuperscript{th} time unit and then decrease in the number of vehicles from 400\textsuperscript{th} to 600\textsuperscript{th} time unit. In this section, we evaluate the performance of DRLD-SP performance by making abrupt changes to the environment. For every 100\textsuperscript{th} time unit, we reduce the vehicular density to 50\%. After 25 time units, it will abruptly change back to the initial pattern. In Fig. \ref{fig:abruptdelay} and Fig. \ref{fig:abruptfairness}, we validate the impact of abrupt changes and the effect on network performance in terms of delay observed by the vehicles and fair deployment of edge resources, respectively. The results show that even with abrupt changes our proposed DRLD-SP framework is effective due to its ability of recording experiences in the replay memory module (Fig. \ref{fig:Model}) which helps to retrain the critic network parameters for better performance in accordance with the changes in the environment.
\subsubsection{Comparison with baseline dynamic frameworks}
\begin{figure}[htbp]
	\centering
	\includegraphics[width=2.8in,height=2.1in]{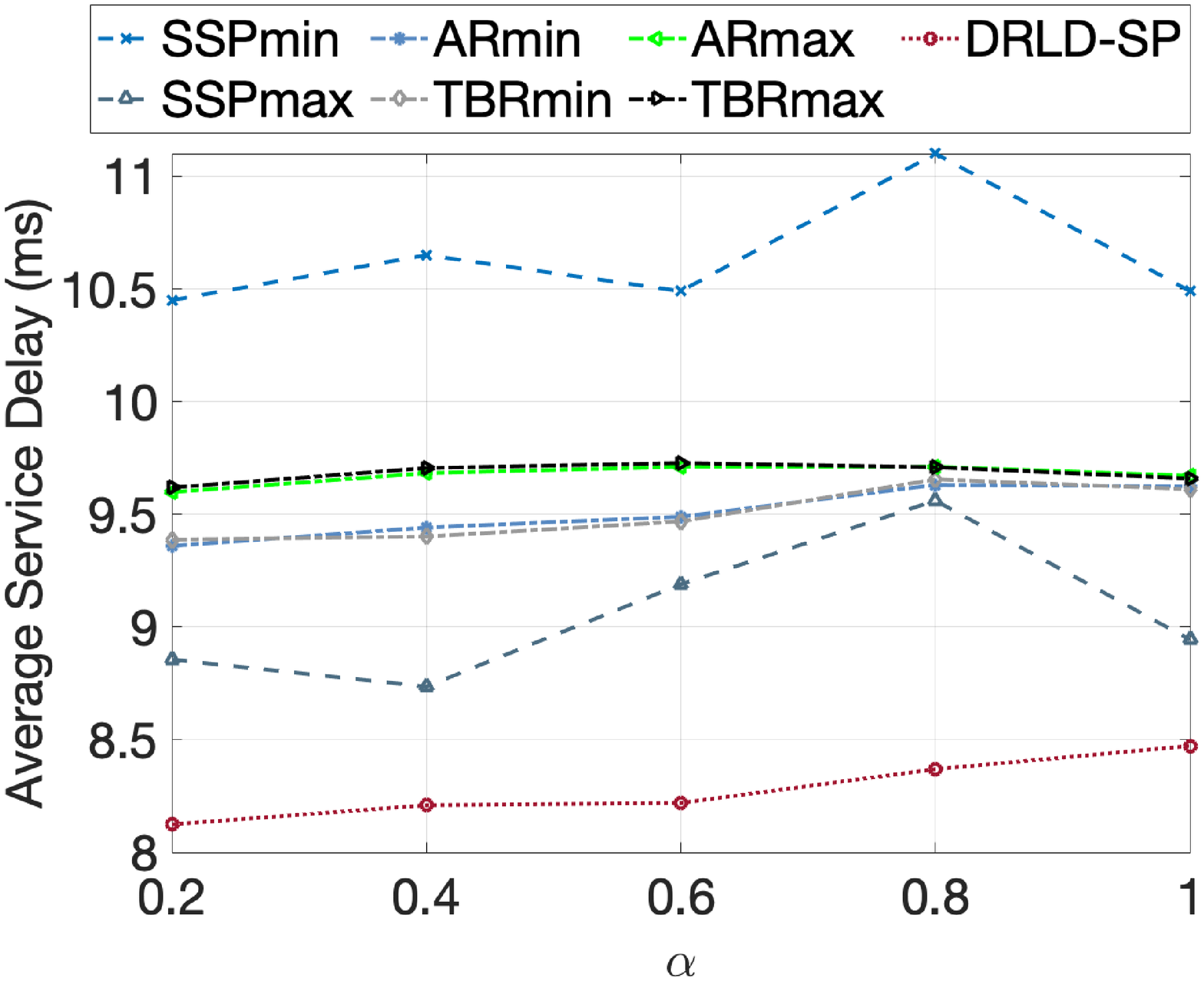}  
	\caption{Average service delay}
	\label{fig:compalldelay}
\end{figure}

\begin{figure}[htbp]
	\centering
	\includegraphics[width=2.8in,height=2.1in]{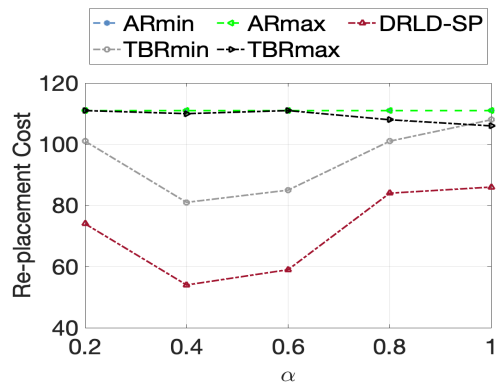}  
	\caption{Re-placement Cost}
	\label{fig:cost}
\end{figure}

We further compare the performance of our proposed DRLD-SP algorithm with the baseline mobility-aware dynamic algorithms AR and TBR (for min and max service instance scenarios). Fig. \ref{fig:compalldelay} and Fig. \ref{fig:cost} compare the average service delay and re-placement cost for different values of $\alpha$, the parameter denoting the relative importance of resource usage vs. delay, respectively. Fig. \ref{fig:compalldelay} shows that DRLD-SP achieves lowest service delay outperforming the static and dynamic schemes. This is because, $AR_{min}$ and $AR_{max}$ in reality doesn't check for the need and dynamicity of the network. It simply finds new optimal solutions in every iteration which affects its performance. Whereas, using a fixed threshold value (in $TBR_{min}$ or $TBR_{max}$) to trigger reoptimization based on previous values of delay, may not work well always as the delay requirements could vary over a wide range. In addition, AR and TBR are mobility-aware dynamic mechanisms that do not consider the dynamicity in terms of service requests and varying (increasing/decreasing) user demands resulting in poor performance. This also degrades service satisfaction for $AR_{min}$ and $TBR_{min}$, as observed in Fig. \ref{fig:ServiceSatisfaction}. On the other hand, using two instances results in resource wastage as observed in Fig. \ref{fig:SInstance}.\par 
Fig. \ref{fig:cost} depicts re-placement cost (migration cost) for different values of $\alpha$. As observed from the figure, the AR and TBR schemes incur higher re-placement cost compared to the proposed DRLD-SP scheme. In $TBR_{max}$, the re-placement cost is higher when compared to $TBR_{min}$ as the latter has not much choices for migration due to resource scarcity. More re-placements of services mean more disconnections and reconnections among vehicles and edge nodes. These frequent service interruptions will affect network performance with additional overheads. In addition, when the value of $\alpha$ is low or high, one of the factors- resource usage or delay- becomes more significant and the re-placement cost becomes high. When the value of $\alpha$ is around 0.5, the delay and resource usage are balanced and the replacement cost is low.    
\section{Conclusion}
\label{sec:conclude}
In this paper, we addressed the problem of dynamic service placement in IoV networks. We developed a deep reinforcement learning-based framework for continual learning of the environment to capture the dynamicity of vehicles, increasing service demands and varying request-types. We formulated the optimization problem to minimize the maximum edge resource usage and service delay. For the decision making, the DRLD-SP agent uses an optimization problem as the actor network and a value-function to critic the quality of the decision taken by the actor network. We evaluated our framework by simulating a virtual traffic scenario of a realistic IoV network using SUMO. We carried out an extensive set of experiments to demonstrate the superiority of our DRLD-SP framework over other static and dynamic placement methods in terms of several important metrics such as delay, resource usage, fairness, and migration cost.

\balance


\bibliographystyle{IEEEtran}
\bibliography{IEEEabrv,References1}

\begin{thebibliography}{10}
\providecommand{\url}[1]{#1}
\csname url@samestyle\endcsname
\providecommand{\newblock}{\relax}
\providecommand{\bibinfo}[2]{#2}
\providecommand{\BIBentrySTDinterwordspacing}{\spaceskip=0pt\relax}
\providecommand{\BIBentryALTinterwordstretchfactor}{4}
\providecommand{\BIBentryALTinterwordspacing}{\spaceskip=\fontdimen2\font plus
\BIBentryALTinterwordstretchfactor\fontdimen3\font minus
  \fontdimen4\font\relax}
\providecommand{\BIBforeignlanguage}[2]{{%
\expandafter\ifx\csname l@#1\endcsname\relax
\typeout{** WARNING: IEEEtran.bst: No hyphenation pattern has been}%
\typeout{** loaded for the language `#1'. Using the pattern for}%
\typeout{** the default language instead.}%
\else
\language=\csname l@#1\endcsname
\fi
#2}}
\providecommand{\BIBdecl}{\relax}
\BIBdecl

\bibitem{IoV5G}
C.~R. {Storck} and F.~{Duarte-Figueiredo}, ``A survey of {5G} technology
  evolution, standards, and infrastructure associated with
  vehicle-to-everything communications by internet of vehicles,'' \emph{IEEE
  Access}, vol.~8, pp. 117\,593--117\,614, 2020.

\bibitem{slicing}
3GPP, \emph{Study on management and orchestration of network slicing for next
  generation network (Release 15)}.\hskip 1em plus 0.5em minus 0.4em\relax
  Document TR 28.801, 2018.

\bibitem{ITUSlice}
M.~Series, \emph{{IMT} Vision--Framework and overall objectives of the future
  development of {IMT} for 2020 and beyond}.\hskip 1em plus 0.5em minus
  0.4em\relax Recommendation ITU, 2015, vol. 2083.

\bibitem{MECSurvey}
P.~{Porambage}, J.~{Okwuibe}, M.~{Liyanage}, M.~{Ylianttila}, and T.~{Taleb},
  ``Survey on multi-access edge computing for internet of things realization,''
  \emph{IEEE Communications Surveys Tutorials}, vol.~20, no.~4, pp. 2961--2991,
  2018.

\bibitem{etsiMEC}
ETSI, \emph{Multi-Access Edge Computing {(MEC)}; Study on {MEC} Support for
  {V2X} Use Cases}.\hskip 1em plus 0.5em minus 0.4em\relax European
  Telecommunications Standards Institute, ETSI GR MEC 022 V2.1.1, Sept 2018.

\bibitem{RLforITS}
A.~{Haydari} and Y.~{Yilmaz}, ``Deep reinforcement learning for intelligent
  transportation systems: A survey,'' \emph{IEEE Transactions on Intelligent
  Transportation Systems}, pp. 1--22, 2020.

\bibitem{driving}
B.~R. {Kiran}, I.~{Sobh}, V.~{Talpaert}, P.~{Mannion}, A.~A.~A. {Sallab},
  S.~{Yogamani}, and P.~{Pérez}, ``Deep reinforcement learning for autonomous
  driving: A survey,'' \emph{IEEE Transactions on Intelligent Transportation
  Systems}, pp. 1--18, 2021.

\bibitem{RL1}
G.~{Wang} and F.~{Xu}, ``Regional intelligent resource allocation in mobile
  edge computing based vehicular network,'' \emph{IEEE Access}, vol.~8, pp.
  7173--7182, 2020.

\bibitem{cache3}
G.~{Qiao}, S.~{Leng}, S.~{Maharjan}, Y.~{Zhang}, and N.~{Ansari}, ``Deep
  reinforcement learning for cooperative content caching in vehicular edge
  computing and networks,'' \emph{IEEE Internet of Things Journal}, vol.~7,
  no.~1, pp. 247--257, 2020.

\bibitem{cache2}
L.~T. {Tan} and R.~Q. {Hu}, ``Mobility-aware edge caching and computing in
  vehicle networks: A deep reinforcement learning,'' \emph{IEEE Transactions on
  Vehicular Technology}, vol.~67, no.~11, pp. 10\,190--10\,203, 2018.

\bibitem{R1}
X.~Hou, Z.~Ren, J.~Wang, W.~Cheng, Y.~Ren, K.-C. Chen, and H.~Zhang, ``Reliable
  computation offloading for edge-computing-enabled software-defined iov,''
  \emph{IEEE Internet of Things Journal}, vol.~7, no.~8, pp. 7097--7111, 2020.

\bibitem{R2}
Z.~Ning, K.~Zhang, X.~Wang, L.~Guo, X.~Hu, J.~Huang, B.~Hu, and R.~Y.~K. Kwok,
  ``Intelligent edge computing in internet of vehicles: A joint computation
  offloading and caching solution,'' \emph{IEEE Transactions on Intelligent
  Transportation Systems}, vol.~22, no.~4, 2021.

\bibitem{schedule}
W.~{Zhan}, C.~{Luo}, J.~{Wang}, C.~{Wang}, G.~{Min}, H.~{Duan}, and Q.~{Zhu},
  ``Deep-reinforcement-learning-based offloading scheduling for vehicular edge
  computing,'' \emph{IEEE Internet of Things Journal}, vol.~7, no.~6, pp.
  5449--5465, 2020.

\bibitem{schedule2}
E.~K. Ahmed, Z.~Li, B.~Veeravalli, and S.~Ren, ``Reinforcement learning-enabled
  genetic algorithm for school bus scheduling,'' \emph{Journal of Intelligent
  Transportation Systems}, vol.~0, no.~0, pp. 1--19, 2020.

\bibitem{navigation}
A.~{Anwar} and A.~{Raychowdhury}, ``Autonomous navigation via deep
  reinforcement learning for resource constraint edge nodes using transfer
  learning,'' \emph{IEEE Access}, vol.~8, pp. 26\,549--26\,560, 2020.

\bibitem{Anum}
A.~{Talpur} and M.~{Gurusamy}, ``Reinforcement learning-based dynamic service
  placement in vehicular networks,'' in \emph{(to appear) IEEE 93rd Vehicular
  Technology Conference (VTC2021-Spring)}, 2021.

\bibitem{Edge-enabledV2X}
A.~{Moubayed}, A.~{Shami}, P.~{Heidari}, A.~{Larabi}, and R.~{Brunner},
  ``Edge-enabled {V2X} service placement for intelligent transportation
  systems,'' \emph{IEEE Transactions on Mobile Computing}, pp. 1--1, 2020.

\bibitem{multicomponent-V2X}
I.~Shaer, A.~Haque, and A.~Shami, ``Multi-component {V2X} applications
  placement in edge computing environment,'' in \emph{ICC 2020 - 2020 IEEE
  International Conference on Communications (ICC)}, 2020, pp. 1--6.

\bibitem{costoptimalplacement}
A.~{Moubayed}, A.~{Shami}, P.~{Heidari}, A.~{Larabi}, and R.~{Brunner},
  ``Cost-optimal {V2X} service placement in distributed cloud/edge
  environment,'' in \emph{2020 16th International Conference on Wireless and
  Mobile Computing, Networking and Communications (WiMob)}, 2020, pp. 1--6.

\bibitem{cloud2}
Y.~{Lin}, J.~{Hu}, B.~{Kar}, and L.~{Yen}, ``Cost minimization with offloading
  to vehicles in two-tier federated edge and vehicular-fog systems,'' in
  \emph{2019 IEEE 90th Vehicular Technology Conference (VTC2019-Fall)}, 2019.

\bibitem{cloud3}
H.~Yao, C.~Bai, D.~Zeng, Q.~Liang, and Y.~Fan, ``Migrate or not? exploring
  virtual machine migration in roadside cloudlet-based vehicular cloud,''
  \emph{Concurrency and Computation: Practice and Experience}, vol.~27, no.~18,
  pp. 5780--5792, 2015.

\bibitem{priority1}
X.~{Yu}, M.~{Guan}, M.~{Liao}, and X.~{Fan}, ``Pre-migration of vehicle to
  network services based on priority in mobile edge computing,'' \emph{IEEE
  Access}, vol.~7, pp. 3722--3730, 2019.

\bibitem{D1latencyaware}
B.~E. Mada, M.~Bagaa, T.~Tale, and H.~Flinck, ``Latency-aware service placement
  and live migrations in 5g and beyond mobile systems,'' in \emph{ICC 2020 -
  2020 IEEE International Conference on Communications (ICC)}, 2020, pp. 1--6.

\bibitem{D2drl}
Y.~Peng, L.~Liu, Y.~Zhou, J.~Shi, and J.~Li, ``Deep reinforcement
  learning-based dynamic service migration in vehicular networks,'' in
  \emph{2019 IEEE Global Communications Conference (GLOBECOM)}, 2019, pp. 1--6.

\bibitem{D3cloud}
T.~Taleb, A.~Ksentini, and P.~A. Frangoudis, ``Follow-me cloud: When cloud
  services follow mobile users,'' \emph{IEEE Transactions on Cloud Computing},
  vol.~7, no.~2, pp. 369--382, 2019.

\bibitem{D4fog}
W.~Bao, D.~Yuan, Z.~Yang, S.~Wang, W.~Li, B.~B. Zhou, and A.~Y. Zomaya,
  ``Follow me fog: Toward seamless handover timing schemes in a fog computing
  environment,'' \emph{IEEE Communications Magazine}, vol.~55, no.~11, pp.
  72--78, 2017.

\bibitem{dynamicRA}
H.~{Liang}, X.~{Zhang}, X.~{Hong}, Z.~{Zhang}, M.~{Li}, G.~{Hu}, and F.~{Hou},
  ``Reinforcement learning enabled dynamic resource allocation in internet of
  vehicles,'' \emph{IEEE Transactions on Industrial Informatics}, pp. 1--1,
  2020.

\bibitem{containerization}
J.~Zhang, X.~Zhou, T.~Ge, X.~Wang, and T.~Hwang, ``Joint task scheduling and
  containerizing for efficient edge computing,'' \emph{IEEE Transactions on
  Parallel and Distributed Systems}, vol.~32, no.~8, pp. 2086--2100, 2021.

\bibitem{etsi5G}
ETSI, \emph{{5G}; Study on scenarios and requirements for next generation
  access technologies}.\hskip 1em plus 0.5em minus 0.4em\relax European
  Telecommunications Standards Institute, ETSI TR 138 913 V15.0.0, 2018.

\bibitem{queueMD1}
B.~Jansson, ``Choosing a good appointment system - a study of queues of the
  type {(D, M, 1)},'' \emph{Operations Research}, vol.~14, no.~2, 1966.

\bibitem{OpenStreetMap}
{OpenStreetMap contributors}, ``{Planet dump retrieved from
  https://planet.osm.org },'' \url{ https://www.openstreetmap.org }, 2017.

\bibitem{jain_index}
R.~Jain, D.~Chiu, and W.~Hawe, \emph{A Quantitative Measure of Fairness and
  Discrimination for Resource Allocation in Shared Computer System}.\hskip 1em
  plus 0.5em minus 0.4em\relax Eastern Research Laboratory, Digital Equipment
  Corporation, 1984.

\end{thebibliography}

\end{document}